\documentclass[prd,preprint,tightenlines,floatfix,showpacs,preprintnumbers,nofootinbib,eqsecnum]{revtex4}
 \usepackage[dvips,final]{graphicx}
  \usepackage{amssymb}
   \usepackage{amsmath}
    \usepackage{amsfonts}
     \usepackage{epsfig}
      \usepackage{bm}

\def\mb{\,\mbox{mb}}

\def\GeV{\,\mbox{GeV}}

\def\Pom{{ I\!\!P}}
 \def\Reg{{ I\!\!R}}

 \newcommand\beq{\begin{equation}}
 
 \newcommand\eeq{\end{equation}}
 \newcommand\beqn{\begin{eqnarray}}
 \newcommand\eeqn{\end{eqnarray}}

\begin{document}

\begin{flushright}
LU TP 13-38\\
USM-TH-322\\
October 2015
\end{flushright}

\title{Diffractive Higgsstrahlung}

\author{Roman Pasechnik}
 \email{Roman.Pasechnik@thep.lu.se}
 \affiliation{Department of Astronomy and Theoretical
 Physics, Lund University, SE 223-62 Lund, Sweden}

\author{Boris Kopeliovich and Irina Potashnikova}
 \affiliation{Departamento de F\'{\i}sica,
Universidad T\'ecnica Federico Santa Mar\'{\i}a; and\\
Centro Cient\'ifico-Tecnol\'ogico de Valpara\'{\i}so,
Avda. Espa\~na 1680, Valpara\'{\i}so, Chile\vspace{1cm}}

\begin{abstract}
\vspace{0.5cm} 
We consider single-diffractive (SD) Higgs production in association
with heavy flavour in proton-proton collisions at the LHC. The main focus 
of our study is a reliable estimate of SD/inclusive ratio, not a precision 
computation of the cross sections. The calculations are performed within 
the framework of the phenomenological dipole approach, which includes 
by default the absorptive corrections, i.e. the gap survival effects 
at the amplitude level. The dominant mechanism is the diffractive 
production of heavy quarks, which radiate a Higgs boson (Higgsstrahlung). 
Although diffractive production of $t$-quarks is grossly suppressed as $1/m_t^2$,
the large Higgs-top coupling compensates this smallness and 
the Higgsstrahlung by $t$-quarks becomes the dominant contribution 
at large Higgs boson transverse momenta. We computed the basic 
observables such as the transverse momentum and rapidity distributions 
of the diffractively produced Higgs boson in association with the bottom and top 
quark pair. Finally, we discuss a potential relevance of the diffractive Higgsstrahlung
in comparison to the Higgsstrahlung off intrinsic heavy flavor at forward 
Higgs boson rapidities.
\end{abstract}


\maketitle

\section{Introduction}

The Higgs boson recently discovered at the LHC \cite{ATLAS,CMS}
appears to be one of the most prominent standard candles for
physics within and beyond the Standard Model (SM) (for more details of
the Higgs physics highlights at the LHC see e.g.
reviews ~\cite{Carena:2002es,Handbook} and references
therein). Most of the SM extensions predict stronger or weaker 
distortions in Higgs boson Yukawa couplings. In
this sense, measurements of the Higgs-heavy quarks couplings become
a very important task of the ongoing Higgs physics studies at the
LHC and serve as one of the major probes for the signals of New Physics.

Phenomenological tests of a number of New Physics scenarios at a TeV
energy scale relies upon our understanding of the underlined QCD
dynamics and backgrounds. The QCD-initiated gluon-gluon fusion 
mechanism is one of the dominant Higgs bosons production modes in 
inclusive $pp$ scattering which has contributed to its discovery at the LHC 
\cite{ATLAS,CMS}. The hard loop induced amplitude has been studied in a wealth 
of theoretical articles so far. The inclusive cross section has been calculated at up 
to next-to-next-to-leading order in QCD \cite{Georgi:1977gs,Djouadi:1991tka,
Dawson:1990zj,Spira:1995rr,Anastasiou:2002yz,Ravindran:2003um} and recently 
up to N$^3$LO level \cite{Ball:2013bra}. Also, the 
QCD soft-gluon re-summation at up to next-to-next-to-leading logarithm approximation 
was performed in Ref.~\cite{Catani:2003zt} and the next-to-leading order factorized 
electroweak corrections were incorporated in Ref.~\cite{Actis:2008ug}. Besides the
standard collinear factorisation approach, inclusive Higgs boson production has 
been studied in the $k_\perp$-factorisation framework in 
Refs.~\cite{Lipatov:2005at,Lipatov:2014mja}. The inclusive associated
production of the Higgs boson and heavy quarks has been thoroughly 
analyzed in the $k_\perp$-factorisation in Ref.~\cite{Lipatov:2009qx}.

The phenomenological studies of inclusive Higgs boson production channels 
typically suffer from large Standard Model backgrounds
and theoretical uncertainties, strongly limiting their potential
for tracking possibly small New Physics effects. As a promising way
out, the exclusive and diffractive Higgs production processes offer new
possibilities to constrain the backgrounds, and open up more
opportunities for New Physics searches (see e.g.
Refs.~\cite{Durham,Heinemeyer:2007tu,Heinemeyer:2010gs,Tasevsky:2013iea}). 
Likewise in inclusive production, the loop-induced gluon-gluon 
fusion $gg\to H$ mechanism is expected to be an important 
Higgs production channel in single diffractive $pp$ scattering as well, 
while this is the only possible mechanism for the central exclusive 
Higgs production \cite{Durham}.

Once the poorly known nonperturbative elements are constrained
by pure SM-driven data sets, they can also be
applied for description of other sets of data potentially sensitive
to New Physics contributions. This way, it would be possible to pin
down and to constrain the yet unknown sources of theoretical
uncertainties purely {\it phenomenologically} to a precision
sufficient for searches of new phenomena at the LHC.
In particular, diffractive production of heavy flavoured particles at forward
rapidities is often considered as one of  the important probes for the
QCD dynamics at large distances, which can be efficiently constrained
by data. 

The understanding of the mechanisms of inelastic diffraction 
came with the pioneering works of Glauber \cite{Glauber}, Feinberg and 
Pomeranchuk \cite{FP56}, Good and Walker \cite{GW} where diffraction 
is conventionally viewed as shadow of inelastic processes. This picture is realised, 
in particular, in the framework of the dipole approach \cite{zkl} where a diffractive 
process looks like elastic scattering of $\bar qq$ dipoles of different sizes, and 
of higher Fock states containing more partons. 

By construction, the phenomenological color dipole approach effectively takes into account 
the major part of the higher-order and soft QCD corrections \cite{nik}. In particular, the dipole model
predictions appear to be very close to the corresponding predictions of the collinear factorisation 
approach at Next-to-Leading Order (NLO) for Drell-Yan production process \cite{Raufeisen:2002zp} 
as well as in heavy flavor production \cite{Kopeliovich:2003cn}. In this sense, the dipole approach
is analogical to the $k_\perp$-factorisation technique (see e.g. Ref.~\cite{Lipatov:2009qx} and 
references therein).

Besides, it provides a prominent way to
study the diffractive factorisation breaking effects due to an interplay between
hard and soft interactions. In addition, the gap survival effects are effectively
taken into account at the amplitude level. Previously, the latter effects have been
successfully studied in the case of forward Abelian radiation of
virtual photons (diffractive Drell-Yan reaction) in Refs.~\cite{Kopeliovich:2006tk,Pasechnik:2011nw}, 
as well as for the more general case of forward gauge bosons production \cite{Pasechnik:2012ac}, 
and in the non-Abelian case of the forward heavy flavour production
\cite{Kopeliovich:2007vs}.
The main ingredient of the dipole formalism is the process-independent universal dipole-target scattering 
cross section. It can thus be determined phenomenologically, for example, from the Deep Inelastic Scattering 
(DIS) data \cite{GBWdip}.
Based on the formalism developed earlier
\cite{Kopeliovich:2007vs,Pasechnik:2012ac}, in this work we employ the colour
dipole approach specifically for the inclusive and, for the first time, single diffractive
Higgs boson production in association with a heavy quark pair in
proton-proton collisions at the LHC. 

Since the Higgs boson-quark couplings in the SM are proportional to the
quark masses, a significant contribution to the Higgs production at
forward rapidities comes from the Higgsstrahlung process off heavy
quarks (predominantly, off bottom $b$ and top $t$ quarks) in the
proton sea. Furthermore, in this work we do not take into
consideration the inclusive and diffractive Higgsstrahlung mechanism off 
the intrinsic heavy flavours, which was previously studied in
Refs.~\cite{Brodsky:2007yz,Brodsky:2006wb}. Here we consider
diffractive Higgsstrahlung off heavy quarks produced via the
perturbative gluon-gluon fusion mechanism which plays a role as 
the major background component for the forward diffractive Higgsstrahlung off
intrinsic heavy flavour. A relative smallness of the production modes over 
the intrinsic ones at forward rapidities would be an important message 
for future forward diffractive Higgs production studies.

The paper is organized as follows. Section II is devoted to a
discussion of inclusive Higgsstrahlung off heavy quarks in the dipole framework 
in the large Higgs boson transverse momentum limit which is then used in
derivation of the SD-to-inclusive ratio. The corresponding amplitude in momentum 
space is derived in Appendix A. The SD Higgsstrahlung process has been thoroughly analysed 
both analytically and numerically within the dipole picture in Section III. In particular, 
the SD-to-inclusive ratio has been obtained in analytic form and applied to get an estimate 
for the SD Higgsstrahlung cross section. Such a ration has been verified against 
the SD-to-inclusive ratio for beauty production at CDF Tevatron \cite{Affolder:1999hm} 
(for more details, see Ref.~\cite{Kopeliovich:2007vs}). Finally, basic conclusions 
are made in Section IV.

\section{Inclusive Higgsstrahlung in the dipole picture}

Inclusive production of heavy quarks in association with the Higgs boson
at the leading order has been studied earlier in detail in the framework of 
the $k_T$-factorisation approach in Refs.~\cite{Lipatov:2005at,Lipatov:2009qx}. 
In this Section, we investigate the corresponding process in the dipole framework.

The basic strategy here is to derive an approximated dipole formula for the inclusive cross section 
valid at large Higgs boson transverse momenta and then to use it in derivation of analytic expression
for the SD-to-inclusive ratio. The latter can then be employed beyond the high-$p_T$ approximation
and would provide an important answer about a relative smallness of the SD Higgsstrahlung component,
which is the basic goal of this work.

\begin{figure*}[!h]
 \centerline{\includegraphics[width=0.7\textwidth]{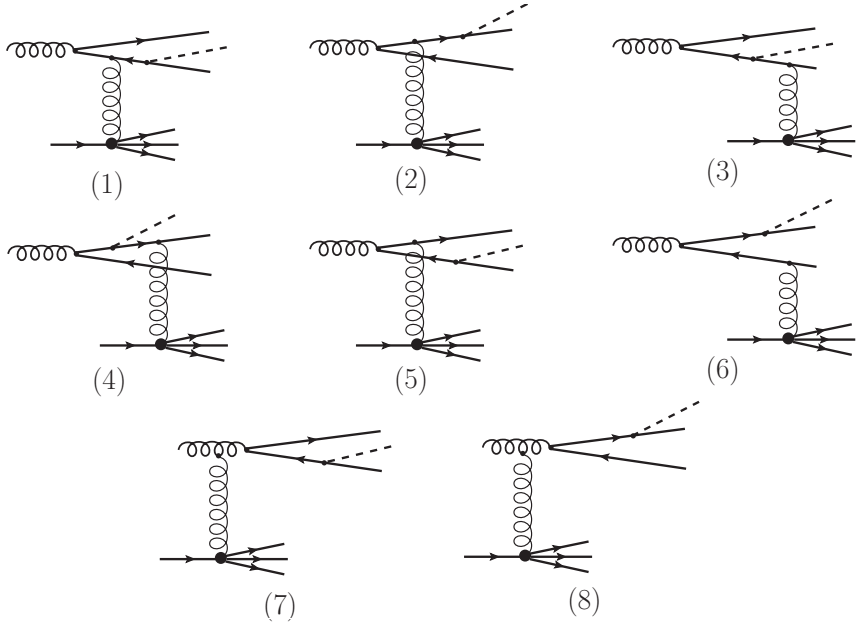}}
   \caption{
\small Leading-order gluon-initiated contributions to the inclusive $\bar Q Q H$ system production 
in the gluon-proton scattering in the proton rest frame.}
 \label{fig:QQh-inclusive}
\end{figure*}

The amplitude of the inclusive Higgsstrahlung process in gluon-proton scattering
by means of single gluon exchange in the $t$ channel
\begin{eqnarray}
G_a + p \to Q\bar QH + X \,, \qquad Q=c,b,t \,,
\end{eqnarray}
where $G_a$ is the initial gluon in colour state $a$, is described in
Born approximation by the set of eight diagrams shown in Fig.~\ref{fig:QQh-inclusive}.

Note, the Born-level diagrams are shown in Fig.~\ref{fig:QQh-inclusive} only 
for illustration. By the virtue of the color dipole framework, the inclusion of 
the universal dipole cross section phenomenologically generalises the Born 
diagrams and effectively resums the lower gluonic ladder diagrams at small x 
to all orders, similarly to the $k_\perp$-factorisation technique. The upper 
partonic ladder will be accounted via DGLAP evolution of the gluonic density 
at the NLO.

The kinematics and a detailed derivation of the corresponding amplitude in impact parameter representation
are presented in Appendix~\ref{appendix}. In order to derive the dipole formula for the corresponding process, 
one should switch to the impact parameter representation performing 2D Fourier transform over the relative 
transverse momentum between $Q$ and $\bar Q$, $\vec \varkappa$, total transverse momentum of the $\bar QQH$ system,
$\vec k_\perp$, and the transverse momentum of the Higgs boson, $\vec \kappa$, defined in the gluon-target c.m. frame in the limit $\kappa \gg k_\perp$ and $\alpha_3\ll 1$. 
The latter differs from the standard Higgs boson transverse momentum defined in proton-target c.m. frame.
Corresponding amplitudes in the impact parameter space are
\begin{eqnarray} \nonumber
A^{\mu\bar\mu}_{a}\equiv \int \frac{d^2 k_\perp}{(2\pi)^2}\frac{d^2 \varkappa}{(2\pi)^2}\frac{d^2 \kappa}{(2\pi)^2}\,
B^{\mu\bar\mu}_{a}\,e^{-i \vec k_\perp\vec s -i \vec \varkappa\vec r -i \vec \kappa\vec\rho} = 
3\sum_{d=1}^{N_c^2-1}\,{\xi_Q^\mu}^\dagger  \Big\{ \tau_a\tau_d\, \hat {\cal T}^{(d)}_1 + \tau_d\tau_a\, \hat {\cal T}^{(d)}_2 \Big\} 
\tilde{\xi}_{\bar Q}^{\bar \mu}\,,
\end{eqnarray}
in terms of the impact parameter dependent amplitudes
\begin{eqnarray} \nonumber
\hat {\cal T}^{(d)}_1(\vec s,\vec r, \vec \rho) &=& \hat \Psi_{1}(\alpha,\alpha_3;\vec r,\vec \rho)\,
\Big[ \hat \gamma^{(d)}(\vec{s} - \alpha \vec r) - \hat \gamma^{(d)}\Big(\vec{s} - \alpha \vec r - 
\frac{\alpha_3}{\bar \alpha}(\vec{\rho}+\alpha \vec{r})\Big) \Big]  \,, \nonumber \\
\hat {\cal T}^{(d)}_2(\vec s,\vec r, \vec \rho) &=& \hat \Psi_{2}(\alpha,\alpha_3;\vec r,\vec \rho)\,
\Big[ \hat \gamma^{(d)}(\vec{s} + \bar\alpha \vec r) - \hat \gamma^{(d)}\Big(\vec{s} + 
\bar\alpha \vec r -\frac{\alpha_3}{\alpha}(\vec{\rho}-\bar \alpha \vec{r})\Big) \Big] \,.\nonumber
\end{eqnarray}
Here, we introduced short-hand notations for the respective production of the wave functions for given scales
\begin{eqnarray}
\hat \Psi_{1}(\alpha,\alpha_3;\vec r,\vec \rho) &\equiv & \frac{\alpha_3}{\bar \alpha}\,\hat \Phi_{\bar QQ}(\vec r, m_Q)\,
\hat \Phi_{QH}(\vec{\rho}+\alpha\vec{r},\tau) \,, \label{amps-TF-1} \\
\hat \Psi_{2}(\alpha,\alpha_3;\vec r,\vec \rho) &\equiv & \frac{\alpha_3}{\alpha}\,\hat \Phi_{\bar QH}(-\vec{\rho}+\bar \alpha\vec{r},\tau)\,
\hat \Phi_{\bar QQ}(\vec r,m_Q) \,, \label{amps-TF-2}
\end{eqnarray}
where $\tau$ is the hard scale determined in Eq.~(\ref{tau}), 
$\tau_a$ are the standard $SU(N_c)$ generators related to the Gell-Mann matrices as $\lambda_a=\tau_a/2$, and 
the gluon-target interaction amplitude $\hat \gamma^{(d)}(\vec{s})$ is an operator in colour and coordinate space of 
the target quarks defined as \cite{Kopeliovich:2001ee}
\begin{eqnarray}
\hat \gamma^{(d)}(\vec{s})\equiv \frac{\sqrt{\alpha_s}}{\sqrt{6}}\,\int \frac{d^2 k_\perp}{(2\pi)^2}\, 
\frac{\hat F^{(d)}_{Gp\to X}(\vec k_\perp,\{X\})}{\vec k_\perp^2+m_g^2}\; e^{-i \vec k_\perp\vec s} = 
\sum_j \tau^{(j)}_a \chi(\vec{s} - \vec b_j)\,. \label{gam}
\end{eqnarray}
Here, $\chi(\vec{s} - \vec b_j)$ is the interaction amplitude of projectile heavy quark with 
$j$-th constituent valence quark in the target proton, $\vec{s}$ is the transverse distance between projectile heavy quark
and the center of gravity of the target, $\vec b_j$ the transverse distance between $j$-th constituent 
valence quark in the target and the center of gravity of the target. 

The $G_a\to Q\bar Q$ and $Q/\bar Q\to Q/\bar Q+H$ distribution amplitudes in impact parameter 
representation read
\begin{eqnarray} \label{QQ-wf}
\hat \Phi_{\bar QQ}(\vec r,\epsilon) &\equiv & \int \frac{d^2\varkappa}{(2\pi)^2}\,
 \hat \Theta_{\bar QQ}(\vec \varkappa,\epsilon)\, e^{-i\vec \varkappa\vec r} \\
&=& \frac{\sqrt{\alpha_s}}{(2\pi)\sqrt{2}} \, 
\Big\{m_Q(\vec e\cdot \vec \sigma)+i(1-2\alpha)(\vec\sigma\cdot \vec n)(\vec e\cdot \vec\nabla_r)-
(\vec e\times \vec n)\cdot \vec\nabla_r \Big\}\,K_0(\epsilon\, r) \,, \nonumber  \\
\hat \Phi_{QH}(\vec{\rho},\varepsilon) &\equiv & 
\int \frac{d^2\kappa}{(2\pi)^2}\,\hat \Theta_{QH}(\vec \kappa,\varepsilon)\, 
e^{-i\vec \kappa\vec \rho} = \frac{m_Q}{(2\pi)\sqrt{3\pi}v}\, 
\Big\{ 2m_Q\,\vec{\sigma}\cdot \vec{n} + \, \vec{\sigma}\cdot \vec\nabla_\rho \Big\}
\,K_0(\varepsilon\,\rho) \,, \label{Qh-wf} \\
\hat \Theta_{\bar QQ}(\vec \varkappa,\epsilon) & \equiv & \frac{\sqrt{\alpha_s}}{\sqrt{2}}\,
\frac{\hat{U}(\vec\varkappa)}{\vec \varkappa^2 + \epsilon^2} \,, \qquad
\hat \Theta_{QH}(\vec \kappa,\varepsilon) \equiv  \frac{m_Q}{\sqrt{3\pi}v}\,
\frac{\hat{V}(\gamma)}{\vec  \kappa^2 + \varepsilon^2 }\,, \label{QQh-mom-wfs}
\end{eqnarray}
respectively, where $\hat{V}(\gamma)$ ($\gamma=\alpha_3/\bar\alpha$ or $\alpha_3/\alpha$) and $\hat{U}(\vec\varkappa)$ are defined in Eq.~(\ref{UV}), 
$r\equiv |\vec r|$ and $\rho \equiv |\vec \rho|$. When taking square of the total inclusive $G_a+p\to Q\bar QH+X$ amplitude 
in impact parameter representation
\begin{eqnarray}
\overline{ |A|^2 }(\vec r_1, \vec \rho_1;\vec r_2, \vec \rho_2)\equiv \int d^2 s\, d\{X\}\sum_{\lambda_{*},a,\mu,\bar\mu}
\langle A^{\mu\bar\mu}_{a}(\vec r_1, \vec \rho_1) \big(A^{\mu\bar\mu}_{a}\big)^\dagger(\vec r_2, \vec \rho_2)\rangle_{|3q\rangle_1}
\end{eqnarray}
one implicitly performs an averaging over colour indices and polarisation $\lambda_*$ 
of the incoming projectile gluon $G_a$ in the $G_a\to Q\bar Q$ and $Q/\bar Q\to Q/\bar Q+H$ distribution amplitudes 
as well as over valence quarks and their relative coordinates in the target proton $|3q\rangle_1$. 
Besides, one uses the general properties of the 2-spinors
\begin{eqnarray}
\sum_{\mu,\bar\mu}\tilde{\xi}_{\bar Q}^{\bar \mu}
\big({\xi_Q^\mu}^\dagger\big)^*= \hat{1} \,, \qquad
\sum_{\mu,\bar\mu}\big({\xi_Q^\mu}^\dagger \hat a  \tilde{\xi}_{\bar Q}^{\bar \mu} \big)^*\, 
\big({\xi_Q^\mu}^\dagger \hat b  \tilde{\xi}_{\bar Q}^{\bar \mu}\big)  = {\rm Tr}\big(\hat a^\dagger \hat b\big) \,.
\end{eqnarray}
Then, squaring the operator $\hat \gamma^{(d)}$ and then averaging it over 
quark positions and quantum numbers the initial nucleon wave function $|3q\rangle_1$ 
and summing over the final $\{X\}$ leads to
\begin{eqnarray}
 \int d\{X\} \langle i|\hat{\gamma}_a(\vec s_k)\hat{\gamma}^\dagger_{a'}(\vec s_l)| i 
\rangle_{|3q\rangle_1} = \frac{1}{8}\, \delta_{aa'}\, S(\vec s_k,\vec s_l)\,, \label{sqgam}
\end{eqnarray}
where the colour averaging procedure
\begin{eqnarray}
\langle i | \tau_a^{(j)}\cdot \tau_{a'}^{(j')} | i \rangle_{|3q\rangle_1} = 
\left\{
     \begin{array}{lr}
       \frac{1}{6}\delta_{aa'} & : j=j'\\
       -\frac{1}{12}\delta_{aa'} & : j\not=j'
     \end{array}
   \right.
\end{eqnarray}
has been performed, and $S(\vec s_k,\vec s_l)$ is the scalar function given by
\begin{eqnarray*}
S(\vec s_k,\vec s_l) &\equiv& \frac{2}{9} \int d\{b\} \Big[ \sum_{j=1}^3 
\chi(\vec{s}_k - \vec{b}_j)\chi(\vec{s}_l - \vec{b}_j) \\ &-& 
\frac12 \sum_{j\not=j'} \chi(\vec{s}_k - \vec{b}_j)\chi(\vec{s}_l - \vec{b}_{j'}) \Big]\,|\Phi_{|3q\rangle_1}(\{\vec b\})|\,.
\end{eqnarray*}
in terms of the quark-target scattering amplitude, $\chi(\vec R)$, and the proton wave function, $\Phi_{|3q\rangle_1}\{\vec b\}$. 
This function is directly related to the universal $q\bar q$ dipole cross section 
known from phenomenology as follows
\beq
\sigma_{\bar q q}(\vec{s}_1-\vec{s}_2)\equiv \int d^2 s \Big[ S(\vec s+\vec{s}_1,\vec s+\vec{s}_1)+
S(\vec s+\vec{s}_2,\vec s+\vec{s}_2)-2S(\vec s+\vec{s}_1,\vec s+\vec{s}_2)  \Big]\,.   \label{sint}
\eeq

The universal dipole cross section $\sigma_{\bar qq}$ implicitly depends on energy. Although being universal, 
it cannot be calculated reliably from the first principles, but is known from phenomenology.
A popular simple GBW ansatz for the saturated shape of the dipole cross section with 
$x_2$-dependent parameters fitted to the HERA hard DIS data at small $x$ \cite{GBWdip}
\begin{eqnarray}
\sigma_{\bar q q}(\vec R)=\bar \sigma_0\Big[ 1 - e^{-\vec R^2/\bar R_0^2(x_2)} \Big] \label{GBW}
\end{eqnarray}
is sufficient for our purposes here since the typical hard scale of inclusive Higgsstrahlung 
is large, i.e. $\mu\sim M \gg m_g$, and thus all the incident dipole sizes are small compared to the
hadron scale (the latter is not true for diffraction, see below). In this case, due to the colour transparency property
it suffices to use
\begin{eqnarray}
\sigma_{\bar qq}(\vec R) \simeq \frac{\bar \sigma_0}{\bar R_0^2(x_2)} \; \vec R^2 \,, \qquad R\ll \bar R_0(x_2) \,, \label{CT}
\end{eqnarray}
to the first approximation. The GBW fits provide 
\beq
\bar \sigma_0=23.03\; \mathrm{mb}\,, \qquad \bar R_0(x)=0.4\,\mathrm{fm}\times(x/x_0)^{0.144}\,, \qquad x_0=3.04\times10^{-4} \,.
\label{GBWpar}
\eeq

Following to the above scheme one obtains the amplitude squared $\overline{|A|^2}$
in an analytic form as a linear combination of the dipole cross sections for different dipole separations,
with coefficients given by colour structure and distribution amplitudes. The dipole formula for the differential cross section 
of the $G_a+p \to \bar QQH + X$ process then reads
\begin{eqnarray}
 \frac{d\sigma(Gp \to \bar QQH + X)}{d\alpha d\ln \alpha_3} = \int d^2rd^2\rho\; \overline{|A|^2}(\vec r, \vec \rho) \,,
\label{Gp-CS}
\end{eqnarray}
Here, the amplitude squared in the general form
\begin{eqnarray} \nonumber
\overline{|A|^2}(\vec r, \vec \rho)=\frac38(\Psi_{1}\Psi_{2}^\dagger + \Psi_{2}\Psi_{1}^\dagger)\, \sigma_{\rm eff}(\vec r, \vec R_1, -\vec R_2)
+ \, 6\big(\Psi_{1}\Psi_{1}^\dagger\,\sigma_{\bar qq}(\vec R_1) + \Psi_{2}\Psi_{2}^\dagger\,\sigma_{\bar qq}(\vec R_2) \big)\,,
\end{eqnarray}
where
\begin{eqnarray} \nonumber
&& \sigma_{\rm eff}(\vec l_1,\vec l_2,\vec l_3) \equiv \sigma_{\bar qq}(\vec l_1) - \sigma_{\bar qq}(\vec l_1 + \vec l_3) - 
\sigma_{\bar qq}(\vec l_1 + \vec l_2) + \sigma_{\bar qq}(\vec l_1 + \vec l_2 + \vec l_3) \,,  \nonumber \\
&& \vec R_1 \equiv \frac{\alpha_3}{\bar \alpha}\,(\vec \rho + \alpha \vec r)\,, \quad 
\vec R_2 \equiv \frac{\alpha_3}{\alpha}\,(\vec \rho - \bar \alpha \vec r)\,.  \label{WQP}
\end{eqnarray}

Transition to the hadron level is usually performed as follows (see e.g. Ref.~\cite{Kopeliovich:2001ee})
\begin{eqnarray} \label{dipole-f}
&&\frac{d\sigma_{\rm incl}}{dY\,d\alpha\,d\ln \alpha_3} = G(x_1,\mu^2)\,\frac{d\sigma(Gp \to \bar QQH + X)}{d\alpha d\ln \alpha_3}\,,
\end{eqnarray}
where the projectile gluon distribution in the incoming proton is treated via the collinear factorisation technique, 
$Y$ is the rapidity of the $\bar QQH$ system, and
\begin{eqnarray}
G(x_1,\mu^2)\equiv x_1g(x_1,\mu^2) \,, \qquad x_{1,2}=\frac{M}{\sqrt{s}}\,e^{\pm Y}\,. \label{x1x2}
\end{eqnarray}
are the collinear gluon density at the hard scale $\mu^2\simeq M^2$ being the invariant mass
of the $Q\bar QH$ system defined in Eq.~(\ref{invMass}) and the momentum fractions of 
the projectile and $t$-channel gluon, respectively.
\begin{figure*}[!ht]
 \centerline{\includegraphics[width=0.5\textwidth]{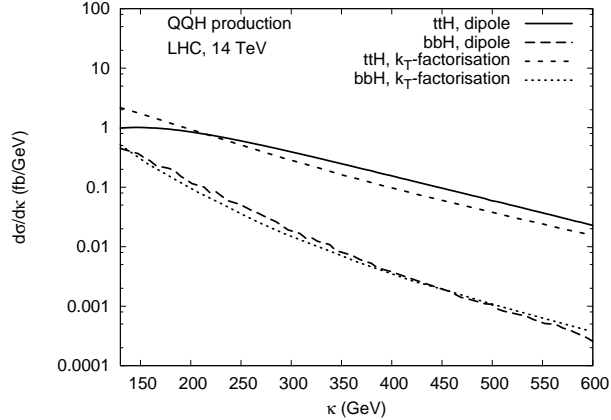}}
   \caption{ \small The differential Higgs boson transverse momentum $\kappa$ distribution 
   of the inclusive Higgsstrahlung obtained via the asymptotic dipole formula (\ref{incl-CS}) 
   valid in the limit of small $k_\perp \ll \varkappa,\,\kappa$ and $\alpha_3\ll 1$ 
   in comparison to the exact result of $k_\perp$-factorisation from Ref.~\cite{Lipatov:2009qx}.}
 \label{fig:incl-QQh-pt}
\end{figure*}

In order to obtain the Higgsstrahlung cross section differential in relative transverse momenta $\vec \kappa$ and $\vec \varkappa$,
one can directly use the asymptotic amplitudes (\ref{T1sS}) and (\ref{T1sS}) in the limit
of small $k_\perp \ll \varkappa,\,\kappa$. At the level of cross section, this asymptotics is equivalent to taking the first
quadratic term in the dipole cross section (\ref{CT}). In this case, we obtain 
the fully differential inclusive Higgsstrahlung cross section which reads
\begin{eqnarray}\label{incl-CS}
 &&\frac{d\sigma_{\rm incl}}{d\Omega} \simeq
 \frac{1}{(2\pi)^4}\,\frac{3\bar \sigma_0}{4\bar R_0^2(x_2)}\,
 g(x_1,\mu^2)\; \mathrm{Tr}\Big[ 8\vec{\hat{{\cal K}}}_1\cdot \vec{\hat{{\cal K}}}^\dagger_1 + 
 8\vec{\hat{{\cal K}}}_2\cdot \vec{\hat{{\cal K}}}^\dagger_2 -
 \vec{\hat{{\cal K}}}_1\cdot \vec{\hat{{\cal K}}}^\dagger_2 - 
 \vec{\hat{{\cal K}}}_2\cdot \vec{\hat{{\cal K}}}^\dagger_1 \Big] \,,
\end{eqnarray}
where $\vec{\hat{{\cal K}}}_{1,2}$ are defined in terms of momentum-space wave functions (\ref{QQh-mom-wfs}) as
\begin{eqnarray}
\vec{\hat{{\cal K}}}_1 &\equiv & - \frac{\alpha_3}{\bar \alpha}\, 
\hat \Theta_{\bar QQ}(\vec \varkappa - \alpha \vec \kappa, m_Q)\,
 \frac{\partial}{\partial \vec z}\Big[\hat \Theta_{QH}(\vec z,\tau)\Big]_{\vec z = \vec\kappa} \,, \label{K1} \\
\vec{\hat{{\cal K}}}_2 &\equiv & \frac{\alpha_3}{\alpha}\,
\frac{\partial}{\partial \vec z}\Big[\hat \Theta_{QH}(\vec z,\tau)\Big]_{\vec z = - \vec\kappa}\, 
\hat \Theta_{\bar QQ}(\vec \varkappa + \bar \alpha \vec \kappa, m_Q) \,, \label{K2}
\end{eqnarray} 
and
\beq
 d\Omega\equiv dx_1\,d\alpha\,d\ln\alpha_3\,d^2\varkappa\,d^2\kappa  \label{dOm}
\eeq 
is the element of the phase space volume associated with the produced 
system $\bar QQH$. The remaining momentum integrals can then
be numerically evaluated over a given phase space volume specific to a given measurement 
(see below).

In Fig.~\ref{fig:incl-QQh-pt} the approximated dipole formula result 
for the inclusive Higgsstrahlung cross section differential in Higgs boson transverse momentum 
$\kappa$ (\ref{incl-CS}) is compared to the corresponding exact calculation in the $k_\perp$-factorisation 
approach of Ref.~\cite{Lipatov:2009qx}. Remind, the asymptotic dipole formula 
Eq.~(\ref{incl-CS}) is obtained in the collinear projectile gluon and soft target 
gluon $k_\perp \ll \varkappa,\,\kappa$ approximations, as well as for $\alpha_3\ll 1$. 
In this case, the final Higgs boson transverse momentum is entirely generated by a recoil
against heavy quarks in the final state. Besides, the Higgs boson is assumed to take only a relatively
small fraction of the quark momenta. A comparison with the exact result 
of Ref.~\cite{Lipatov:2009qx} shows that the asymptotic Higgs boson 
spectrum (\ref{incl-CS}) generated by purely final state kinematics 
dominates the total cross section at large Higgs boson transverse momenta $\kappa \gtrsim m_H$ 
and approaches the exact result both in shape and normalisation. At lower transverse 
momenta, however, we notice a missing strength due to the omitted diagrams as well as 
a potentially large role of the non-Gaussian tail in primordial gluon transverse momenta distribution.
The latter should be accounted for by the use of unintegrated gluon distribution functions 
as was done in Ref.~\cite{Lipatov:2009qx}.

On the other hand, the simplified dipole formula (\ref{dipole-f}) with a
collinear starting PDF (\ref{x1x2}) and the quadratic approximation in the 
dipole cross section (\ref{CT}) will enable us to calculate the SD-to-inclusive 
ratio in a fully analytic form which does not depend on higher order QCD corrections and 
on projectile gluon evolution and is given only in terms of parameters of the universal dipole 
cross section (see below). As will be discussed below, the ratio is not sensitive to the high-$p_T$ 
approximation we adopted in the analysis of the absolute cross sections as well as 
to the short-distance corrections to the $gg \to Q\bar QH$ subprocess. It therefore can be 
applied to the conventional NNLO+NNLL $Q\bar QH$ inclusive cross sections and/or those obtained in 
the $k_\perp$ factorisation approach known from the literature in order to get a good first estimate 
of the diffractive Higgsstrahlung cross section. 

\section{Single diffractive Higgsstrahlung in the dipole picture}

When it comes to diffraction, the QCD factorisation in hadronic collisions is 
severely broken by the interplay of soft and hard QCD interactions and by the
absorptive effects \cite{Kopeliovich:2006tk,Pasechnik:2011nw}. While the former 
mechanism, which is the leading twist, is frequently missed, the latter effect
is modelled by the gap survival probability factor, which is usually applied to correct the 
factorisation-based results. A successful alternative to the factorisation-based parton model, 
the colour dipole
description \cite{zkl} goes beyond the QCD factorisation and naturally 
accounts for the hard-soft QCD dynamics interplay, and for the absorptive effects
at the amplitude level (see e.g. Refs.~\cite{Kopeliovich:2007vs,Pasechnik:2012ac}).
The single diffractive $Q\bar Q + H$ production (with the leading 
proton and a rapidity gap) is not yet available in the literature, and this Section
is devoted to the corresponding analysis in the dipole approach.

\subsection{Diffractive amplitude}

In order to derive the single diffractive Higgsstrahlung amplitude in impact parameter
representation we refer to the corresponding framework previously developed for
diffractive gluon radiation and diffractive DIS processes in Refs.~\cite{GBWdip,Raufeisen:2002zp,kst1}
and we adopt similar notations in what follows. Similarly to the inclusive case considered above,
we are interested in single-diffractive Higgsstrahlung at high energies when the Higgs boson takes 
relatively small fraction of the heavy quark momentum since 
the well-known wave functions for $G\to Q\bar Q$ 
and $Q(\bar Q)\to Q(\bar Q) + H$ are factorized out in the impact 
parameter space. The latter situation thus enables us to employ
the dipole approach in this first study of the diffractive Higgsstrahlung 
process.
\begin{figure*}[!h]
 \centerline{\includegraphics[width=0.7\textwidth]{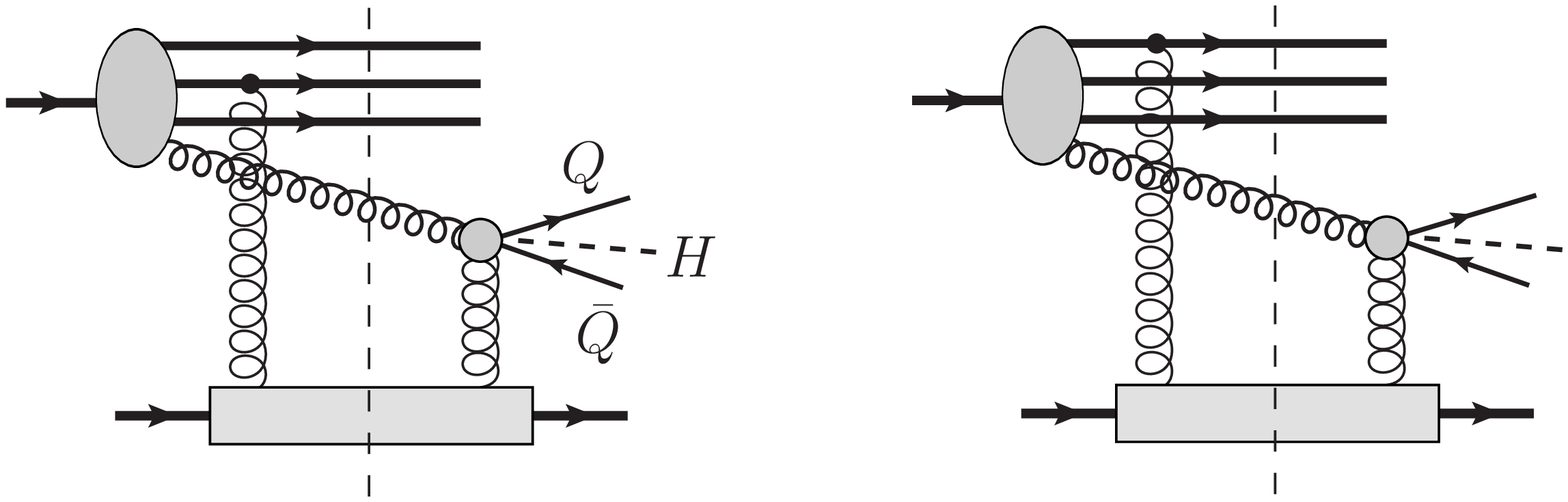}}
   \caption{
\small The dominating gluon-initiated contributions to the single diffractive $\bar Q Q + H$ production in $pp$ 
collisions. The hard $G_a+p\to \bar QQH+X$ subprocess via single gluon exchange where all possible 
di-gluon couplings to $Q\bar Q H$ system are resummed and denoted by filled grey circle and explicitly 
described in Fig.~\ref{fig:QQh-inclusive}. The unitarity cut between the $t$-channel ``active'' (rightmost gluon) and 
``screening'' (leftmost gluon) exchanges is shown by vertical dashed line. In actual calculations, we adopt the
Good-Walker picture of diffraction \cite{GW} (see main text).}
 \label{fig:QQh-SD}
\end{figure*}

For illustration, the dominating parton-level gluon-initiated graphs are shown in Fig.~\ref{fig:QQh-SD} where we account 
only for the diagrams where the ``active'' gluon is coupled to the hard $Q\bar Q+H$ system, 
while the soft ``screening'' gluon couples to a spectator parton at a large impact distance. 
Other diagrams where both ``active'' and ``screening'' gluons couple to partons at small 
relative distances are the higher twist ones and thus are strongly suppressed by extra powers of the hard
scale $\mu^2\sim M^2$ (see e.g. Refs.~\cite{Kopeliovich:2007vs,Pasechnik:2012ac}). 
This becomes even more obvious in the colour dipole framework due to colour transparency 
 \cite{zkl} making the medium more transparent for smaller dipoles. 
 
In what follows, we do not explicitly calculate the Feynman graphs in Fig.~\ref{fig:QQh-SD} but instead 
adopt the Good-Walker picture of diffraction \cite{GW} where a diffractive scattering amplitude
is proportional to a difference between elastic scatterings of different Fock states off the target in the
target rest frame. To this end, applying the generalized optical theorem in the high energy 
limit with a cut between the ``screening'' and ``active'' $t$-channel gluons 
as illustrated in Fig.~\ref{fig:QQh-SD}, we write
\begin{eqnarray}
\hat{A}^a_{\rm SD}(\vec r_1,\vec r_2,\vec r_3)=\frac{i}{2} \sum_{Y^*} \langle {\rm in} |(A^{\mu\bar\mu}_a)^\dagger | Y^* \rangle  
\langle Y^* | A_{\rm scr}(\vec r_1,\vec r_2,\vec r_3) | {\rm in}' \rangle 
\label{D-amp},
\end{eqnarray}
where $A_{\rm scr}$ is the ``screening gluon'' exchange amplitude between a constituent (valence) projectile 
quark and the target, $A^{\mu\bar\mu}_a$ is the ``active gluon'' exchange amplitude between colour-singlet 
$G_a(\bar QQ)H$ system and the target found earlier. In the above equation, 
summation goes through all the intermediate states $\{Y^*\}$ except the projectile gluon $G_a$, and 
$a$ is the colour index of the projectile gluon $G_a$. The latter gluon can be assumed to be decoherent in colour 
w.r.t. valence quarks in the incoming proton wave function and thus its colour should be summed up independently 
of the ``screening'' $A_{\rm scr}$ amplitude at the level of cross section. 

Indeed, the projectile hard gluon before its splitting to $Q\bar QH$ system in the colour field of the target undergoes 
multiple radiation steps $g\to gg$ populating the forward rapidity domain with gluon 
radiation with momenta below the hard scale of the process $p_{\perp}^{rad}<\mu$. 
The radiated gluons should then be resummed and can be taken into account by 
using the corresponding gluon PDF similarly to the inclusive case considered above.
Since the hard gluon experiences many splittings on its way (e.g. radiates many gluons and quarks) before 
it gives rise to the $\bar QQ+H$ system, naturally its colour gets completely uncorrelated
with the colour  of the parent valence quark, which should be taken into account
in the respective colour averaging procedure. So in the single diffractive amplitude squared 
one effectively sums over the projectile gluon colour as follows
\begin{eqnarray}
\overline{|A_{\rm SD}|^2}=\sum_{a=1}^{N_c^2-1}\langle\hat{A}^a_{\rm SD}(\hat{A}^a_{\rm SD})^\dagger\rangle_{|3q\rangle_1} \,,
\end{eqnarray}
where the averaging over the colours of the constituent quarks in the incoming proton wave function $|3q\rangle$
is implicitly performed according to the rule
\begin{eqnarray*}
\langle \hat A\big(\tau^{(1)}\big)\hat B\big(\tau^{(2)}\big)\hat C\big(\tau^{(3)}\big) \rangle_{|3q\rangle_1}&=&
\frac{1}{6} \Big\{ {\rm Tr}\,\hat A(\tau)\;{\rm Tr}\,\hat B(\tau)\;{\rm Tr}\,\hat C(\tau) + 
{\rm Tr}\Big[ \hat A\big(\tau\big)\hat B\big(\tau\big)\hat C\big(\tau\big) \Big] \\ &+& 
{\rm Tr}\Big[ \hat A\big(\tau\big)\hat C\big(\tau\big)\hat B\big(\tau\big) \Big] - 
{\rm Tr}\,\hat A\big(\tau\big)\,\Big[\hat B\big(\tau\big)\hat C\big(\tau\big) \Big] \\ &-& 
{\rm Tr}\,\hat B\big(\tau\big)\,\Big[\hat A\big(\tau\big)\hat C\big(\tau\big) \Big] - 
{\rm Tr}\,\hat C\big(\tau\big)\,\Big[\hat A\big(\tau\big)\hat B\big(\tau\big) \Big] \Big\} \,,
\end{eqnarray*}
where $\hat A\big(\tau^{(1)}\big)$, $\hat B\big(\tau^{(2)}\big)$, and $\hat C\big(\tau^{(3)}\big)$
are arbitrary functions of $\tau^{(j)}$ matrices corresponding to valence quarks with $j=1,2,3$ in the projectile 
proton, respectively.

It is well-known that the gluons in such a gluonic ladder are predominantly located in a close vicinity of the valence quarks in the so-called 
``gluonic spots'' which have mean size of about $\sim 0.3$ fm \cite{KST-par,spots}. Thus, 
to the first approximation one could neglect the distance between the projectile gluon $G_a\to \bar QQ+H$ 
and the closest constituent quark compared to the typical distances between the constituent quarks $\sim 1$ fm. 
Then, the amplitude of the ``screening gluon'' exchange summed over projectile valence quarks $j=1,2,3$ reads
\begin{eqnarray} \label{AI}
A^{(i)}_{\rm scr}(\vec r_1,\vec r_2,\vec r_3)=\sum_{j\not=i,d'} \tau^{(j)}_{d'}
\big\{\hat{\gamma}_{d'}(\vec r_i) - \hat{\gamma}_{d'}(\vec r_i + \vec r_{ij}) \big\} \,,
\end{eqnarray}
where $\vec r_1$ is the impact parameter of the gluon $G_a$ or the closest constituent quark $q_i$ (the gluon $G_a$
is assumed to belong to one of the three ``gluonic spots'' in the projectile proton), $\vec r_{ij}\equiv \vec r_j - \vec r_i$ 
is the distances of the other two constituent quark $q_j$, $j\not=i$ from the $q_i$ quark, and the matrices 
$\hat{\gamma}_a$ are the operators in coordinate and colour space for the target quarks defined in Eq.~(\ref{gam}).
Due to the colour transparency, the soft amplitude (\ref{AI}) naturally vanishes if all the distances in the projectile
proton disappear, i.e. $\vec r_{ij}\to 0$. Finally, one should sum over contributions from 
the valence quarks (or the corresponding ``spots''), i.e.
\begin{eqnarray}
A_{\rm scr}(\vec r_1,\vec r_2,\vec r_3)=\sum_{i=1}^3 A^{(i)}_{\rm scr} \,,
\end{eqnarray}
which is equivalent to accounting for cyclic permutation of the valence quarks at the amplitude level. 

In what follows, it is convenient to choose the following set of independent variables
\begin{eqnarray}
\{\vec r_1,\;\vec r_2,\;\vec r_3\}\qquad \Rightarrow \qquad \{\vec s,\; \vec r_{12},\; \vec r_{13} \}
\end{eqnarray}
where $\vec s$ is the impact parameter of the projectile gluon $G_a$.
In variance with the inclusive case considered above, the diffractive scattering is very sensitive to the typical hadron 
size in the projectile proton, i.e. large hadron-scale dipoles $|\vec{r}_{ij}|\sim b\sim R_p$, $i\not=j$ ($R_p$ is the mean proton size) 
become important and control the diffractive Higgsstrahlung. In this case the Bjorken variable $x$ is ill defined, and a more appropriate
variable is the $pp$ collisions energy. An energy dependent parametrization of the dipole cross section 
with the same saturated shape as in the GBW case (\ref{GBW})
\begin{eqnarray}
\sigma_{\bar q q}(\vec R,\hat s)=\sigma_0(\hat s)\Big[ 1 - e^{-\vec R^2/R_0^2(\hat s)} \Big] \,, \qquad \label{KST}
\end{eqnarray}
with parameters being functions of the gluon-target c.m. energy squared $\hat s=x_1 s$ ($s$ is $pp$ c.m. energy squared), 
\begin{eqnarray}
 R_0(\hat s)=0.88\,\mathrm{fm}\,(s_0/\hat s)^{0.14}\,, \quad \sigma_0(\hat s)=\sigma_{tot}^{\pi p}(\hat s)
 \Big(1+\frac{3R_0^2(\hat s)}{8\langle r_{ch}^2 \rangle_{\pi}}\Big)\,,
 \label{KST-params}
\end{eqnarray}
was proposed and fitted to the soft hadronic data in Ref.~\cite{KST-par}.
Here, the pion-proton total cross section is parametrized as
\cite{barnett} $\sigma_{tot}^{\pi p}(\hat s)=23.6(\hat s/s_0)^{0.08}$ mb,
$s_0=1000\,\GeV^2$, the mean pion radius squared is \cite{amendolia}
$\langle r_{ch}^2 \rangle_{\pi}=0.44$ fm$^2$.

Following the above scheme one obtains the diffractive Higgsstrahlung amplitude $\hat{A}^a_{\rm SD}$ 
in analytic form as a linear combination of the dipole cross sections for different dipole separations.
As was anticipated, the diffractive amplitude represents the destructive interference effect from 
scattering of dipoles of slightly different sizes and vanishes as $\hat{A}^a_{\rm SD}\propto \alpha_3^2$ 
in the limit $\alpha_3\to 0$. Such an interference results in the interplay between hard and soft fluctuations 
in the diffractive $pp$ amplitude, enhancing the breakdown of diffractive factorisation 
\cite{Kopeliovich:2006tk,Pasechnik:2011nw}.

\subsection{The dipole formula for the cross section}

Note that initial $\langle {\rm in}|$ and intermediate $| Y^* \rangle$ states are composite and contain the projectile
proton wave function of the initial proton $\Psi_i(\vec r_l,x_l)$ and the projectile proton remnant wave function 
$\Psi_f(\vec r_l,x_l)$ as a function of positions $\vec r_i$ and momentum fractions $x_i$ of all the incident partons.
Using the squaring rule for gluon-target interactions given by Eq.~(\ref{sqgam}), integrating over $\vec s$ according to Eq.~(\ref{sint}), 
keeping only the leading (quadratic) terms in small $\rho,r\ll r_{ij}$ and Fourier-transforming the amplitude back to momentum space, 
we obtain explicitly
\begin{eqnarray} \nonumber
\overline{|A_{\rm SD}|^2}&\simeq &\frac{3}{256}\, |\Psi_{in}|^2\,|\Psi_{fin}|^2\, \sum_{i,j=1}^2 \Omega^{ij}_{\rm hard}\, \Omega^{ij}_{\rm soft} \,, \\
\Omega^{ij}_{\rm hard}&=&\mathrm{Tr}\Big[ 8\hat{{\cal K}}_{1,i} \hat{{\cal K}}^\dagger_{1,j} + 
 8\hat{{\cal K}}_{2,i} \hat{{\cal K}}^\dagger_{2,j} -
 \hat{{\cal K}}_{1,i} \hat{{\cal K}}^\dagger_{2,j} - 
 \hat{{\cal K}}_{2,i} \hat{{\cal K}}^\dagger_{1,j} \Big]  \,, \nonumber \\  
\Omega^{ij}_{\rm soft}&=&\Big\{ \big[ 2\nabla_i\sigma_{\bar qq}(\vec r_{12}) + \nabla_i\sigma_{\bar qq}(\vec r_{12}-\vec r_{13}) + 
\nabla_i\sigma_{\bar qq}(\vec r_{13}) \big]\nabla_j\sigma_{\bar qq}(\vec r_{12}) \nonumber \\ \nonumber &+& 
\big[ \nabla_i\sigma_{\bar qq}(\vec r_{12}) + 2\nabla_i\sigma_{\bar qq}(\vec r_{12}-\vec r_{13}) - 
\nabla_i\sigma_{\bar qq}(\vec r_{13}) \big]\nabla_j\sigma_{\bar qq}(\vec r_{12} - \vec r_{13})  \\  &+& 
\big[ \nabla_i\sigma_{\bar qq}(\vec r_{12}) - \nabla_i\sigma_{\bar qq}(\vec r_{12}-\vec r_{13}) + 
2\nabla_i\sigma_{\bar qq}(\vec r_{13}) \big]\nabla_j\sigma_{\bar qq}(\vec r_{13}) \Big\}\,, \label{ASD}
\end{eqnarray}
where $\vec{\hat{{\cal K}}}_{1,2}$ are defined earlier in Eqs.~(\ref{K1}) and (\ref{K2}),
\begin{eqnarray}
\nabla_i\sigma_{\bar qq}(\vec R) = \frac{2\sigma_0(\hat s)}{R_0^2(\hat s)}\, R_i\, e^{-R^2/R_0^2(\hat s)}\,,
\end{eqnarray}
with the saturated form of the dipole cross section in the KST (energy dependent) form (\ref{KST}) and (\ref{KST-params}). 
Equation (\ref{ASD}) corresponds to the
single diffractive $\bar QQH$ production process in the Good-Walker picture \cite{GW}, by construction. We notice that 
the soft ($r_{ij}$-dependent) part of the SD amplitude squared in Eq.~(\ref{ASD}) has been accumulated 
in $\Omega^{ij}_{\rm soft}$ while all the dependence on the hard scales $\rho$ and $r$ is contained in $\Omega^{ij}_{\rm hard}$, while 
the partonic structure of the projectile proton is concentrated in $|\Psi_{in}|^2$. The amplitude above
is normalized in such a way that the cross section of the $\bar QQ+H$ production in forward single-diffractive 
$pp$ scattering reads
\begin{eqnarray}
\frac{d\sigma_{\rm SD}}{d\alpha\,d\ln\alpha_3\, d^2\delta_\perp}\Big|_{\delta_\perp\to 0}=
\frac{1}{(2\pi)^2}\, \int \prod_{i,j} d^2r_i d^2r'_j\,\prod_{k,l,m,n} dx_q^k dx_g^l dx_q^{\prime m} dx_g^{\prime n}\, 
\int d^2 r d^2 \rho\, \overline{|A_{\rm SD}|^2} \,, \label{FSD}
\end{eqnarray}
where $\vec \delta_\perp$ is the transverse momentum 
of the final proton associated with the $t$-channel momentum transfer squared,
$t=-|\delta_\perp|^2$, $x_{q/g}^i$ are the fractional light-cone momenta of the valence/sea 
quarks and gluons. As long as the forward diffractive 
cross section (\ref{FSD}) is known, the total SD cross section can be evaluated as
\begin{eqnarray}
\frac{d\sigma_{\rm SD}}{d\Omega} \simeq \frac{1}{B_{\rm SD}(s)}\,
\frac{d\sigma_{\rm SD}}{d\Omega\,d\delta_{\perp}}\Big|_{\delta_{\perp}\to0} \,, \quad 
B_{\rm SD}(s)\simeq \langle r_{ch}^2 \rangle_p/3+2\alpha'_{\Pom}\ln(s/s_1) \,,\quad s_1=1\,{\rm GeV}^2 \label{slope}
\end{eqnarray}
where $d\Omega$ is the element of the phase space volume defined in Eq.~(\ref{dOm}), 
and $B_{\rm SD}(s)$ is the Regge-parameterised $t$-slope of the 
differential SD cross section (with $\alpha'_{\Pom}=0.25\,\GeV^{-2}$), which is expected 
to be similar to the $t$-slope measured in diffractive DIS. 

The initial proton $\Psi_{in}$ and proton remnant $\Psi_{fin}$ wave functions in Eq.~(\ref{ASD})
encode information about kinematics and probability distributions of individual (incoming and outgoing) partons. 
In the unobservable part, the completeness relation to the wave function of the proton remnant $\Psi_{fin}$ in
the final state reads
\begin{eqnarray}\nonumber
&&\sum_{fin}\Psi_{fin}(\vec{r}_1,\vec{r}_2,\vec{r}_3;\{x_q^{1,2,...}\},\{x_g^{1,2,...}\})
\Psi^*_{fin}(\vec{r}\,'_1,\vec{r}\,'_2,\vec{r}\,'_3;\{{x'}_q^{1,2,...}\},\{{x'}_g^{1,2,...}\})\\
&&\phantom{.......}=\,
\delta\bigl(\vec{r}_1-\vec{r}\,'_1\bigr)\delta(\vec{r}_2-\vec{r}\,'_2)
\delta(\vec{r}_3-\vec{r}\,'_3)\prod_{j}\delta(x_{q/g}^j-{x'}_{q/g}^j) \,.
\end{eqnarray}
In the above formula, $\delta$-functions reflect momentum conservation and will simplify the phase space 
integrations over the unobservable variables in the single diffractive Higgsstrahlung cross section considerably (see below).

The light-cone partonic wave function of the initial proton $\Psi_{in}$ depends on transverse coordinates 
and fractional momenta of all valence and sea quarks and gluons. As was mentioned above, 
we assume that the mean transverse distance between a source valence quark and
the sea quarks or gluons is much smaller than the mean distance between the valence quarks. 
Therefore, the transverse positions of sea quarks and gluons can be identified with the coordinates 
of the valence quarks, and the proton wave function squared $|\Psi_{in}|^2$ can be parametrized as,
\begin{eqnarray}\nonumber
|\Psi_{in}|^2&=&\frac{3a^2}{\pi^2}
e^{-a(r_1^2+r_2^2+r_3^2)}\;{\cal R}\big(x_1,\{x_q^{1,2,...}\},\{x_g^{2,3,...}\}\big)\\
&\times&\delta(\vec{r}_1+\vec{r}_2+\vec{r}_3)\delta\Big(1-x_1-\sum_j x_{q/g}^j\Big), \label{psi}
\end{eqnarray}
where $a\equiv \langle r_{ch}^2 \rangle_p^{-1}$ is the inverse proton mean charge 
radius squared, the variable $x_g^1\equiv x_1$ is defined as the light-cone momentum fraction 
of the hard gluon related to rapidity $Y$ of the produced $\bar QQ+H$ system in Eq.~(\ref{x1x2}); 
${\cal R}$ is a valence/sea (anti)quark distribution function in the projectile proton.
For simplicity, we parametrize the valence part of the proton wave function
in the form of symmetric Gaussian for the spacial quark distributions. Notice 
that ${\cal R}$ distribution has a low (hadronic) scale, so
the constituent quarks, i.e. the valence quarks together with the sea and gluons they generate, 
carry the whole momentum of the proton, by construction.

In the case of diffractive gluon excitations, after integration over the fractional momenta of all partons 
not participating in the hard interaction, we arrive at a single gluon distribution in the proton, probed by 
the heavy system $M$,
\begin{eqnarray}\label{single-quark}
\int \prod_{j} dx_q^j \prod_{k\not=1}dx_g^k\,\delta\Big(1-x_1-\sum_l
x_{q/g}^l\Big)\;{\cal R}\big(x_1,\{x_q^{1,2,...}\},\{x_g^{2,3,...}\}\big)=\Big(\frac{C_A}{C_F}\Big)^2 g(x_1,\mu^2)\,,
\end{eqnarray}
in terms of the PDF of the hard gluon $g(x_1,\mu^2)$ with fractional momentum $x_1$, with a proper colour factor being 
the square of the Casimir factor $C_A/C_F=9/4$, where for $N_c=3$ the factors $C_A=N_c=3$
and $C_F=(N_c^2-1)/2N_c=4/3$ are the strengths of the gluon self-coupling and a gluon coupling to a quark,
respectively. While single diffraction with production of  $Q\bar Q$ and hence $Q\bar QH$ is dominated by 
the gluon-gluon fusion (``production''), in the corresponding inclusive process the ``bremsstrahlung'' contribution 
can also be important but only at very forward rapidities currently unreachable for measurements (for more details, 
see Ref.~\cite{Kopeliovich:2007vs}).

In the single diffractive $\bar QQH$ production cross section the phase space integral of the soft part 
$\Omega^{ij}_{\rm soft}$ over the positions of the valence quarks $\vec r_{1,2,3}$ in the proton wave function
can be taken analytically, i.e.
\begin{eqnarray*}
 && \int d^2r_1d^2r_2d^2r_3\,e^{-a(r_1^2+r_2^2+r_3^2)}
 \delta(\vec{r}_1+\vec{r}_2+\vec{r}_3)\, \Omega^{ij}_{\rm soft} = \frac19\int d^2r_{12}d^2r_{13}
 e^{-\frac{2a}{3}(r_{12}^2+r_{13}^2+\vec{r}_{12}\vec{r}_{13})}\, \Omega^{ij}_{\rm soft} \\
&& \qquad \qquad =\; 4\pi^2\sigma_0^2(\hat s)\,\Lambda(\hat s)\, \delta^{ij} \,,
 \end{eqnarray*}
where the inverse proton mean charge radius squared $a\simeq 1.38$ fm$^{-2}$, the soft factor
\begin{eqnarray} \nonumber 
\Lambda(\hat s) &\equiv & \Big[a \left(a
   R_0^2+1\right)^2 \left(a R_0^2+3\right)^2 \left(a R_0^2+4\right)^2 \left(a
   R_0^2+12\right)^2 \left(a^2 R_0^4+8\, a R_0^2+3\right)^2\Big]^{-1}\, \\ \nonumber
  &\times& \big\{5\, a^{10} R_0^{20}+192\, a^9 R_0^{18}+3058\, a^8
   R_0^{16}+26224\, a^7 R_0^{14}+132803\, a^6 R_0^{12}+409968\, a^5 R_0^{10} \\
   &+& 771368\, a^4 R_0^8+855216\, a^3 R_0^6+509454\, a^2 R_0^4+149040\, a R_0^2+18144\big\}
   \label{Lambda}
\end{eqnarray}
and $R_0=R_0(\hat s),\,\sigma_0=\sigma_0(\hat s)$ are defined in Eq.~(\ref{KST-params}), and $\hat s=x_1 s$.

The differential SD Higgsstrahlung cross section appears to be proportional to 
the differential inclusive cross section found earlier in Eq.~(\ref{incl-CS}), namely,
\begin{eqnarray}
 &&\frac{d\sigma_{\rm SD}}{d\Omega} \simeq
 \frac{4\bar{R}_0^2(x_2)}{3\bar{\sigma}_0}\,F_{\rm S}(x_1,s)\,\frac{d\sigma_{\rm incl}}{d\Omega} \,,
\label{QQh-PR-SD}
\end{eqnarray}
where $d\Omega$ is the element of the phase space volume defined in Eq.~(\ref{dOm}), 
\beq
F_{\rm S}(x_1,s)\equiv \frac{729\, a^2\sigma_0(x_1 s)^2\,\Lambda(x_1 s)}{4096\,\pi^2\, B_{\rm SD}(s)}\,,\qquad x_1 = \frac{M}{\sqrt{s}}\,e^{+Y} \label{SF}
\eeq
is the energy dependent soft factor, $\Lambda=\Lambda(\hat s)$, $B_{\rm SD}=B_{\rm SD}(s)$ 
and $\tau=\tau(\alpha_3)$ are defined in 
Eqs.~(\ref{Lambda}), (\ref{slope}) and (\ref{tau}), respectively. Note, the differential 
cross section Eq.~(\ref{QQh-PR-SD}) is the full expression, 
which includes by default the effects of absorption at the amplitude level
via differences of elastic amplitudes fitted to data, and does not
need any extra survival probability factor. This fact has been
advocated in detail in Ref.~\cite{Pasechnik:2012ac}, and we do not
repeat this discussion here.

\subsection{Diffractive-to-inclusive ratio and diffractive Higgsstrahlung cross section}
\label{ratio}

The SD-to-inclusive ratio accounting for differences in respective phase space volumes $\Omega'$ and $\Omega$ 
take the following simple form
\begin{eqnarray}\label{ratSDincl}
R_{\rm SD/incl}(M^2,x_1,s)\equiv \delta(M^2)\,\frac{d\sigma^{\rm SD}/d\Omega'}{d\sigma^{\rm incl}/d\Omega} \simeq
\frac{4\bar{R}_0^2(x_2)}{3\bar{\sigma}_0}\,F_{\rm S}(x_1,s)\,\delta(M^2,s) \,, \quad x_2=\frac{M^2}{x_1 s}\,,
\end{eqnarray}
where $d\sigma^{\rm SD}/d\Omega$ and $d\sigma^{\rm incl}/d\Omega$ are the diffractive 
and inclusive Higgsstrahlung cross sections found above, respectively; $\bar{R}_0=\bar{R}_0(x_2)$, $\bar{\sigma}_0$ 
and $F_{\rm S}=F_{\rm S}(x_1,s)$ are defined in Eqs.~(\ref{GBWpar}) and (\ref{SF}), respectively, and $\delta(M^2,s)$
is the suppression factor caused by an experimental cut on $\xi\equiv 1-x_F$ variable. The latter factor 
needs a more detailed clarification.

Indeed, in order to compare our results for the SD-to-inclusive ratio to experimental data, we have to
introduce in our calculations the proper experimental cuts. For example, in diffractive ($Z$, $W$ \cite{CDF-WZ}, 
heavy flavour \cite{Affolder:1999hm}, etc) production measurements at CDF Tevatron, $0.03<\xi<0.1$ constraint was adopted 
(see e.g. Ref.~\cite{CDF-WZ}). Since our single-diffractive cross section formula (\ref{QQh-PR-SD}) is differential 
in kinematics of the produced $\bar QQH$ system, but not in kinematics of the entire diffractive $\bar QQH+X$ system, 
and experimental cuts on $Y$-rapidity (or $x_1$) distribution of a produced system are typically unavailable, 
a direct implementation of the $\xi$ cuts into our formalism and direct comparison to the CDF data cannot be 
performed immediately.

A way out of this problem has been earlier proposed in Ref.~\cite{Pasechnik:2012ac}. 
At small $\xi\to 0$ one can instead write the single diffractive cross section 
in the phenomenological triple-Regge form \cite{3R},
 \beqn \label{Regge}
 &&  - \frac{d\sigma_{\rm SD}^{pp}}
{d\xi\,dp_T^2} =
\sqrt{\frac{s_1}{s}}\,
\frac{G_{\Pom\Pom\Reg}}
{\xi^{3/2}}\,
e^{-B_{\Pom\Pom\Reg}p_T^2} +
\frac{G_{3\Pom}}
{\xi}\,
e^{-B^{pp}_{3\Pom}p_T^2}\,, \quad B_{\Pom\Pom i} = R^2_{\Pom\Pom i} -
2\alpha^\prime_\Pom\,\ln\xi \,, \\
&& G_{3\Pom}(0)=G_{\Pom\Pom\Reg}(0)=3.2\,\mb/\GeV^2\,, \quad R^2_{3\Pom}=4.2\,\GeV^{-2}\,, \quad 
R^2_{\Pom\Pom\Reg}=1.7\,\GeV^{-2} \,,
\nonumber
 \eeqn
where $i=\Pom,\Reg$, $s_1=1\,\GeV^2$, $\alpha_\Pom(0)=1$ 
and $\alpha^\prime_\Pom\approx 0.25\GeV^{-2}$ is the slope of the Pomeron trajectory 
(for more details, see Ref.~\cite{Pasechnik:2012ac}). Then an effect of the experimental cuts on $\xi$ 
in the phenomenological cross section (\ref{Regge}) and in our diffractive cross section 
calculated above (\ref{QQh-PR-SD}) should roughly be the same and the suppression 
factor $\delta$ in Eq.~(\ref{ratSDincl}) valid at CDF environment can be estimated as \cite{Pasechnik:2012ac}
\beq 
 \delta(M^2,s)=\frac{\int dp_T^2\int_{0.03}^{0.1}d\xi\,d\sigma^{pp}_{\rm SD}/dp_T^2 d\xi}
 {\int dp_T^2\int_{\xi_{\rm min}}^{\xi_{\rm max}}d\xi\,d\sigma^{pp}_{\rm SD}/dp_T^2 d\xi}\,, \qquad
 \xi_{\rm min}\equiv \frac{M_{X,\rm min}^2}{s} \sim \frac{M^2}{s} \,,
 \label{delta-cut}
 \eeq 
Here $M_{X,\rm min}\simeq M$ is the minimal produced diffractive mass containing 
the $\bar QQH$ system only. The value of $\delta=\delta(M^2,s)$ in Ref.~(\ref{delta}) 
is essentially determined by the experimental cuts on $\xi$ and is not sensitive to
the upper limit in denominator, so we fix it at a realistic value e.g.
$\xi_{\rm max}\sim 0.3$ \cite{Pasechnik:2012ac}. As a result, for the suppression factor due to
$\xi$-cut we have
\beq
\sqrt{s}=14\,{\rm TeV}\,, \qquad \delta \simeq 0.18\,\dots\,0.24 
\qquad {\rm for} \qquad  M=150\,\dots\,500\;{\rm GeV} \,,
\eeq
respectively, and weakly depends on c.m. energy.

It turns out that the ratio (\ref{ratSDincl}) is controlled mainly by
soft interactions, i.e. expressed in terms of the soft parameters only
$\bar{R_0}(x_2),\,R_0(\hat s),\,\bar{\sigma}_0$ and $\sigma_0(\hat s)$. 
A slow dependence of these parameters on the collision energy $s$ and 
the hard scale $M^2$ governs such dependence of 
the diffractive-to-inclusive production ratio analogically to diffractive gauge
bosons production case earlier discussed in Ref.~\cite{Pasechnik:2012ac}. 
A measurement of the $M^2$ dependence of such a ratio would enable 
to constrain the $x$- and $s$-dependence of the saturation scale as an important
probe of the undelined soft QCD dynamics.
\begin{figure*}[!ht]
 \centerline{\includegraphics[width=0.6\textwidth]{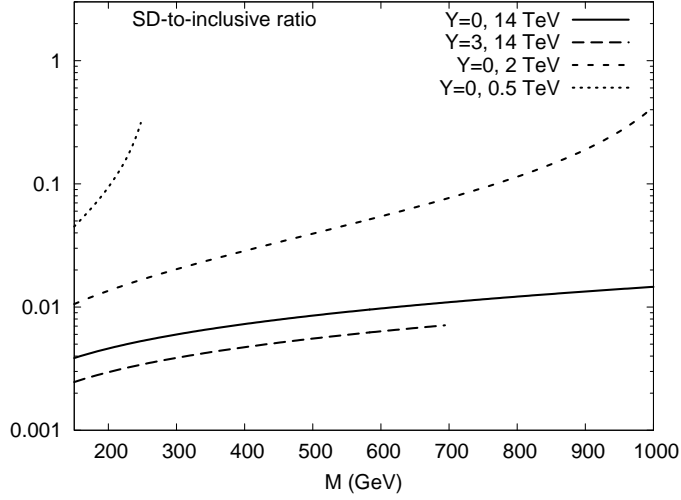}}
   \caption{\small 
  The SD-to-inclusive ratio $R(M)$ as a function of $\bar QQH$ 
  invariant mass $M$ for different c.m. energies $\sqrt{s}=0.5,2,14$ TeV and
  $\bar QQH$ rapidities $Y=0,\,3$.}
 \label{fig:ratio}
\end{figure*}

In Fig.~\ref{fig:ratio} we show the SD-to-inclusive ratio of the cross sections (\ref{ratSDincl}) 
for different c.m. energies $\sqrt{s}=0.5,7,14$ TeV and for two distinct rapidities $Y=0$ and $3$ 
as functions of $\bar QQH$ invariant mass $M$. The effect of additional $\xi$-cuts for 
SD cross section $0.03<\xi<0.1$ is accounted for by means of the multiplicative factor 
$\delta$ defined in Eq.~(\ref{delta-cut}). The considered ratio is found to be consistent 
with the experimentally observed one for diffractive beauty production in 
Ref.~\cite{Affolder:1999hm}. As expected from earlier considerations 
of the diffractive DY \cite{Kopeliovich:2006tk,Pasechnik:2011nw} 
and gauge bosons production \cite{Pasechnik:2012ac} in the dipole framework 
the diffractive factorisation in the SD Higgsstrahlung is broken by transverse motion 
of valence quarks in the projectile proton. The latter effect leads to such unusual 
behavior of the SD-to-inclusive ratio as its growth with the hard scale, $M$, and 
descrease with the c.m. energy, $\sqrt{s}$.
\begin{figure*}[!ht]
\begin{minipage}{0.495\textwidth}
 \centerline{\includegraphics[width=1.0\textwidth]{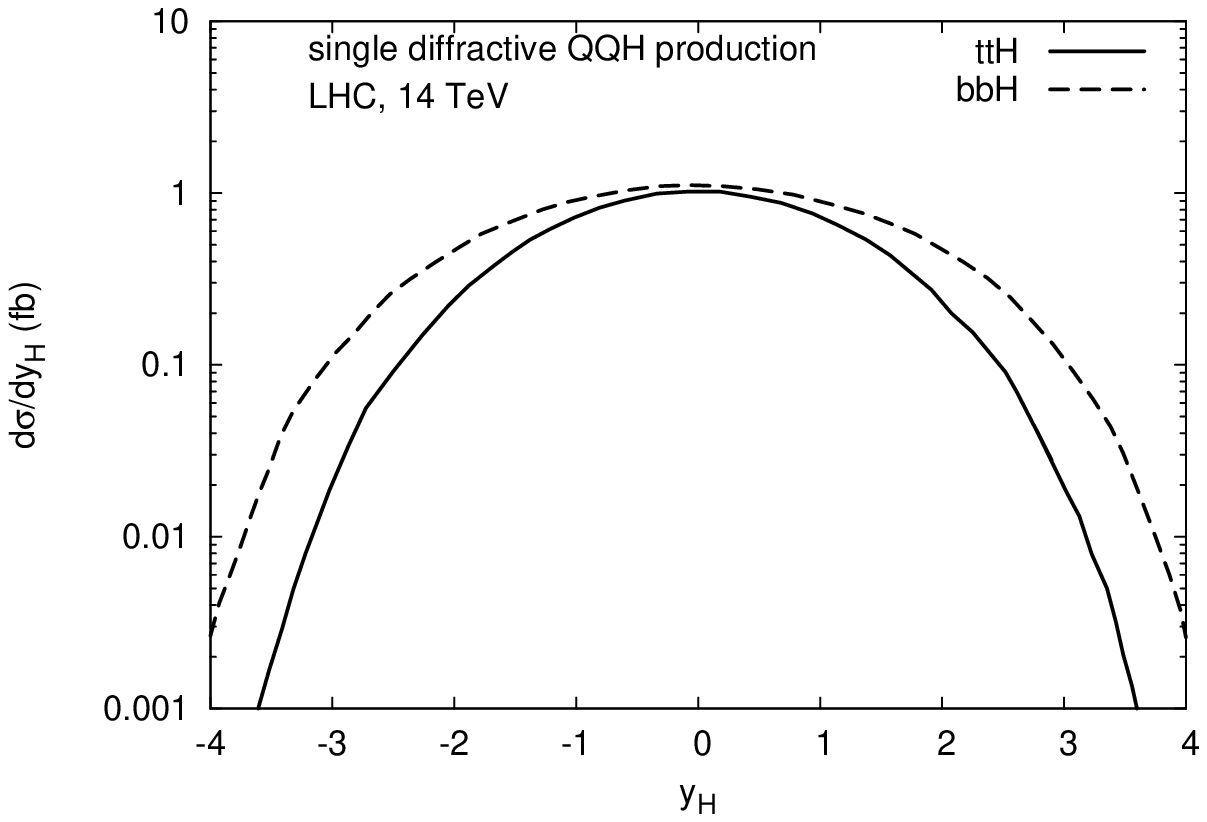}}
\end{minipage}
\begin{minipage}{0.495\textwidth}
 \centerline{\includegraphics[width=1.0\textwidth]{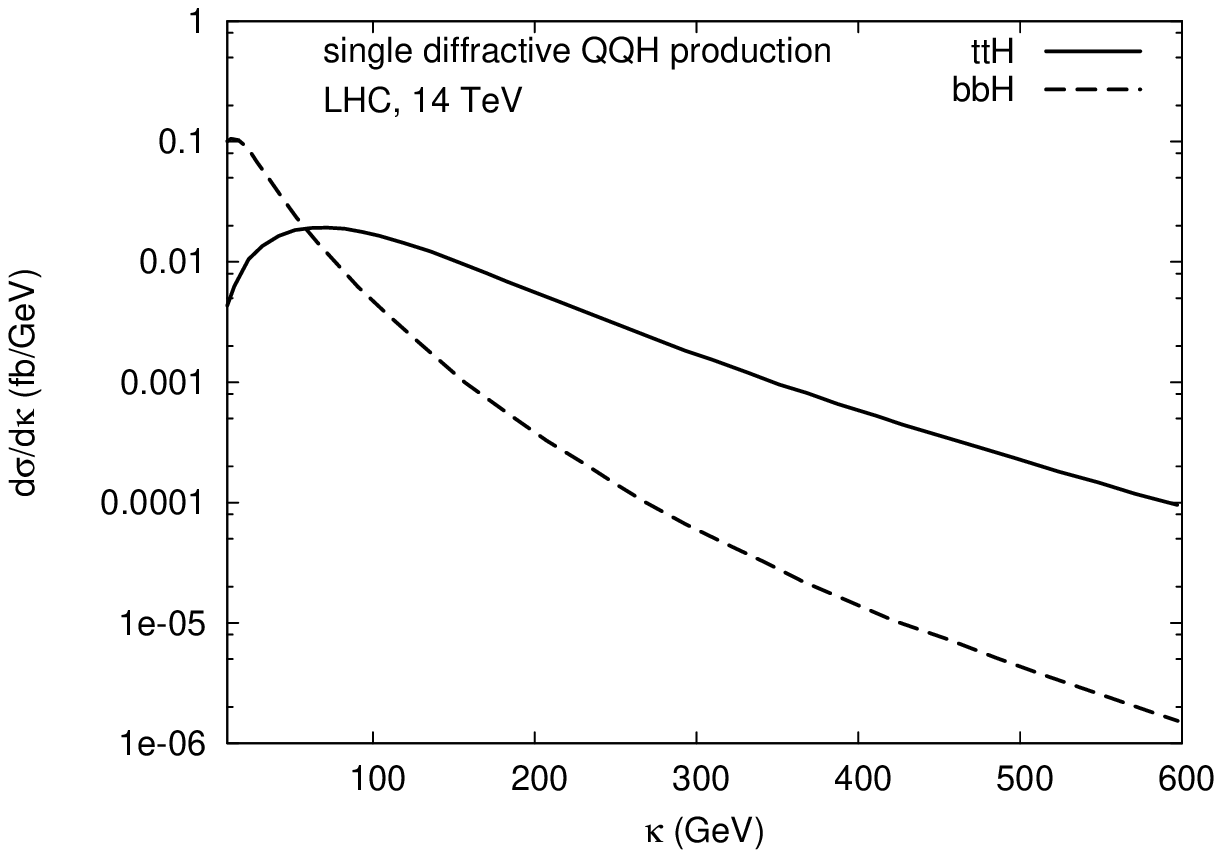}}
\end{minipage}
   \caption{ \small 
  The differential cross sections in Higgs boson rapidity $d\sigma/dy_H$ (left panel) 
  and transverse momentum $d\sigma/d\kappa$ (right panel) of single diffractive 
  Higgsstrahlung off $t\bar{t}$ (solid lines) and $b\bar{b}$ (dashed lines) pairs at the LHC ($\sqrt{s}=14$ TeV). 
  The effect of additional $\xi$-cuts for SD cross section $0.03<\xi<0.1$ 
  is accounted for by means of the multiplicative factor $\delta$ defined 
  in Eq.~(\ref{delta-cut}). The corresponding inclusive cross sections have been
  obtained in the $k_\perp$-factorisation approach with CCFM-evolved unitegrated gluon density following 
  Ref.~\cite{Lipatov:2009qx}.}
 \label{fig:SD-QQh}
\end{figure*}

Due to universality of the SD-to-inclusive ratio (\ref{ratSDincl}) which depends only
on parameters of the dipole cross section it can be applied to the inclusive $Q\bar QH$
production cross section known in the literature to a rather high precision. In order to get 
a reasonable estimate for the SD Higgsstrahlung cross section one can multiply the 
corresponding inclusive cross sections obtained e.g. in the $k_\perp$-factorisation approach 
with CCFM-evolved unitegrated gluon density following Ref.~\cite{Lipatov:2009qx}.
In Fig.~\ref{fig:SD-QQh} we present the resulting curves for the single diffractive 
$pp\to X+(b\bar bH)+p$ (dashed lines) and $pp\to X+(t\bar tH)+p$ (solid lines) cross 
sections differential in Higgs boson rapidity $y_H$ (left panel) and transverse momentum 
$\kappa$ (right panel) at the LHC energy $\sqrt{s}=14$ TeV. At mid-rapidities, the top
and bottom $d\sigma/dy_H$ cross sections turn out to be rather close to each other,
while top contribution strongly dominates over the bottom one at large Higgs boson
transverse momenta $\kappa\gtrsim m_H$. Comparing our results for the production 
$gg\to Q\bar Q H$ mode with the results for the diffractive Higgsstrahlung off the intrinsic 
heavy  flavor from Ref.~\cite{Brodsky:2006wb} we conclude that the
intrinsic contribution to the diffractive Higgs production becomes important 
at rapidities $y_H>3.5$ and should be taken into account.
Of course, a relation of the experimental acceptances for Higgs boson and heavy quark 
decay products with actual phase space bounds on producing Higgs boson and 
heavy quarks is the matter of a dedicated Monte-Carlo 
detector-level simulation (see e.g. Refs.~\cite{Heinemeyer:2007tu,
Heinemeyer:2010gs,Tasevsky:2013iea}) which can be done in the future
if necessary.

\section{Summary}

Here we presented the first calculation of the single diffractive (SD) Higgsstrahlung process off heavy (top and bottom) quarks.
We compute the SD-to-inclusive ratio within the light-cone dipole approach. For this purpose, we estimate the transverse momentum distribution of 
the inclusive and SD cross sections in the dipole framework at large Higgs boson transverse momenta $(\kappa > m_H)$ and observe that they are
proportional to each other. Thus, the considered high-$p_\perp$ limit enables us to extract the SD-to-inclusive ratio in a simple analytic form which then 
could be used beyond the adopted approximations. The ratio between them takes a simple analytic form and depends only on parameters of the dipole 
cross section. By using the naive GBW parameterisation we estimate the numerical accuracy of this ratio for not too large $Q\bar QH$ invariant masses to be
within a factor of two. Such a theoretical uncertainty accounts for typical uncertainties in the choice of available parameterisations for the dipole cross 
section (or unintegrated gluon densities). So by applying the ratio given by Eq.~(\ref{ratSDincl}) to the inclusive Higgsstrahlung $HQ\bar Q$ cross sections 
obtained elsewhere one would get a reasonable estimate for the SD Higgsstrahlung cross section which can be used in practice.

For the SD case, we numerically evaluated the corresponding differential cross sections
in transverse momentum of the Higgs boson and relative transverse momentum of heavy quarks. Similarly to other 
hard diffractive processes \cite{Kopeliovich:2006tk,Pasechnik:2011nw,Pasechnik:2012ac,Kopeliovich:2007vs},
breakdown of QCD factorization leads to rather mild scale dependence of the cross section, $1/m_Q^2$
(compare with $1/m_Q^4$ in diffractive DIS). Such a leading twist behaviour is confirmed by 
the comparison of data on diffractive production of charm and beauty \cite{Kopeliovich:2007vs}.
Radiation of the Higgs boson enhances the relative contribution of heavy flavours at high transverse momenta. 
In Eq.~(\ref{QQh-PR-SD}) one also observes a peculiar feature of similarity of the slopes of differential 
in $\kappa$ inclusive and diffractive cross sections (while the slope for beauty is larger than for top as 
is seen in Fig.~\ref{fig:SD-QQh}) (right panel). This could be anticipated, since the main fraction
of the transferred momentum originates from the short distance interaction, which is the same 
in inclusive and diffractive processes.

The same scale-dependence, $1/m_Q^2$, for diffractive and inclusive cross sections leads 
to flavour independence of their ratio. This is an apparent manifestation of diffractive factorization breaking,
indeed, such a ratio in DIS is steeply falling with $m_Q$. The Higgs couplings cancel in the ratio.
The SD-to-inclusive ratio is similar to that for heavy quark production, which was calculated 
in \cite{Kopeliovich:2007vs} in good agreement with experimental data from the Tevatron.  

Another interesting feature of SD-to-inclusive Higgstrahlung ratio, which can be observed in Fig.~\ref{fig:SD-QQh}, 
is its falling energy- and rising $M$-dependence, where $M$ is the invariant mass of the produced 
$\bar QQH$ system. This is similar to what was found for diffractive Drell-Yan process 
\cite{Kopeliovich:2006tk,Pasechnik:2011nw} and has the same origin: breakdown of QCD factorization 
and the saturated form of the dipole cross section. Of course, the corresponding SD $Q\bar QH$ cross 
sections by themselves are rather small and may not be measurable with the current LHC instrumentation. 
On the other hand, these correspond to a background for the SD Higgsstrahlung off intrinsic heavy flavor 
\cite{Brodsky:2006wb}. The intrinsic mode increases relative to the background and becomes important 
at large rapidities $y_H>3.5$.

Within the dipole model the effects of absorption are included by default in a most natural way, quantum-mechanically. 
They are contained in the parametrization of the dipole cross section fitted to experimental data, and the dipole formula 
for diffractive scattering is self-contained and does not require any extra factors. These corrections are 
accounted for in our calculations at the amplitude level, while most of the existing calculations of the absorption 
effects have been calculated so far probabilistically. The differences might be large, although are difficult to quantify 
at present. Unfortunately, calculation of the central double-diffractive particle production within the dipole approach 
is still a big challenge. Besides, the probabilistic methods are process and kinematics dependent.
Nevertheless, a detailed comparison of dipole model predictions with probabilistic estimates
for the gap survival is a doable problem which is planned for further studies.

\appendix

\section{Inclusive Higgsstrahlung amplitude in momentum space}
\label{appendix}

Explicitly, the inclusive Higgsstrahlung amplitude in gluon-proton collision is described by the set of eight diagrams shown in Fig.~\ref{fig:QQh-inclusive} and reads
\begin{eqnarray} \label{amp-tot}
B^{\mu\bar\mu}_{a}=\frac{i\alpha_s\,m_Q}{\sqrt{4\pi}v}\sum_{l=1}^8
\sum_{d=1}^{N_c^2-1} T_{ad}^{(l)}\, \frac{\hat F^{(d)}_{Gp\to X}(\vec k_\perp,\{X\})}
{\vec k_\perp^2+m_g^2}\, \frac{ {\xi_Q^\mu}^\dagger
\hat{\Gamma_l}\tilde{\xi}_{\bar Q}^{\bar \mu}}{D_l} \,,
\end{eqnarray}
where $v\simeq 246$ GeV is the standard Higgs vacuum expectation value entering the Yukawa couplings in the Standard Model \cite{PDG}, 
$\alpha_s$ is the QCD coupling, $D_l$ are the propagators defined below, $N_c=3$ is the number of colours, $\xi$ are the heavy quark two-component spinors, 
$\tilde{\xi}_{\bar Q}^{\bar\mu}=i\sigma_y (\xi_{\bar Q}^{\bar\mu})^*$, $m_g$ is the effective gluon mass which serves as 
an infrared regulator, $\{X\}$ is the set of variables describing the final state $X$, $\hat F^{(d)}_{Gp\to X}(\vec k_\perp,\{X\})$ is 
the amplitude of the $t$-channel gluon interaction with the proton target $p$ in the target rest frame which determines 
the unintegrated gluon density ${\cal F}$ as follows \cite{Kopeliovich:2001ee}
\begin{eqnarray}
&& \int d\{X\} \sum_{d=1}^{N_c^2-1} |F^{(d)}_{Gp\to X}(\vec k_\perp,\{X\})|^2 = 4\pi {\cal F}(\vec k_\perp^2,x_2) \,, \qquad x_2 = \frac{M^2}{x_1\,s}\,, \\ 
&& M^2 = \frac{m_Q^2+\vec k_1^2}{\alpha_1} + \frac{m_Q^2+\vec k_2^2}{\alpha_2} 
+ \frac{m_H^2+\vec k_3^2}{\alpha_3} \,, \quad \vec k_\perp = \sum_{i=1}^3 \vec k_i\,, \quad \sum_{i=1}^3 \alpha_i = 1 \,,
\end{eqnarray}
where 
\[ \alpha_i = \frac{k_i^+}{k^+} \,, \qquad k^+ = \sum_{i=1}^3 k_i^+ \,, \]
$M$ is the invariant mass of the produced $\bar QQH$ system, $\vec k_{1,2,3}$ and $\alpha_{1,2,3}$ are the transverse 
momenta and fractions of the initial light-cone momentum of the projectile gluon carried by the produced heavy quarks $\bar Q$, $Q$ 
(with mass $m_Q$) and Higgs boson $H$ (with mass $m_H\simeq 126$ GeV), respectively, and $s$ is the Mandelstam variable being 
the total energy of the $pp$ collisons in the $pp$ c.m.s. frame. The colour matrices $T^{(l)}_{ad}(ij)$ for $l$-th diagram in 
Eq.~(\ref{amp-tot}) act in the colour space of the $Q\bar Q$ and have indices $i,\,j$ corresponding to the $Q$ and $\bar Q$, respectively,
\begin{eqnarray} \nonumber
&& T_{ad}^{(1)}=T_{ad}^{(3)}=T_{ad}^{(6)}=\tau_a\tau_d \,, \quad T_{ad}^{(2)}=T_{ad}^{(4)}=T_{ad}^{(5)}=\tau_d\tau_a \,, \\
&& T_{ad}^{(7)}=T_{ad}^{(8)}=i\sum_e^{N_c^2-1} f_{ade}\tau_e = \tau_a\tau_d -\tau_d\tau_a \,,
\end{eqnarray}
where $\tau_a$ are the standard $SU(N_c)$ generators related to the Gell-Mann matrices as $\lambda_a=\tau_a/2$. 

In what follows it would be instructive to introduce the quark momentum fraction 
relative to the $Q\bar Q$ pair
\begin{eqnarray}
\alpha=\frac{k_1^+}{q^+} \,, \qquad q^+ = k_1^+ + k_2^+\,,
\end{eqnarray}
such that
\begin{eqnarray}
\alpha_1=\alpha \bar\alpha_3 \,, \qquad \alpha_2= \bar\alpha \bar\alpha_3 \,.
\end{eqnarray}
Then, the relative transverse momenta between the heavy quark and antiquark, $\varkappa$, and 
between the radiated Higgs boson and $Q\bar Q$ pair, $\kappa$, are 
\begin{eqnarray}
\vec \varkappa=\bar\alpha \vec k_1 - \alpha \vec k_2\,, \qquad 
\vec \kappa = \bar \alpha_3 \vec k_3 - \alpha_3 (\vec k_1 + \vec k_2)\,, \label{k12-kappa}
\end{eqnarray}
respectively, serve as convenient phase space variables of the considering reaction such that the element
of the phase space is
\begin{eqnarray}
d\Omega\propto d\alpha\; d\ln\alpha_3\; d^2\varkappa\; d^2\kappa\,.
\end{eqnarray}
The incident transverse momenta $\vec k_i$ are then defined as
\begin{eqnarray*}
&& \vec k_1 = \vec \varkappa - \alpha \big[\vec \kappa - \bar\alpha_3 \vec k_\perp \big] \,, \\
&& \vec k_2 = - \vec \varkappa - \bar\alpha \big[\vec \kappa - \bar\alpha_3 \vec k_\perp \big] \,, \\
&& \vec k_3 = \vec \kappa + \alpha_3 \vec k_\perp \,,
\end{eqnarray*}
such that in the limit of small $\alpha_3 k_\perp \ll \kappa$, the variable $\vec \kappa$ has the meaning 
of the transverse momentum of the Higgs boson, while
the transverse momentum of the $Q\bar Q$ pair is given by
\begin{eqnarray}
\vec k_{Q\bar Q}\equiv \vec k_1 + \vec k_2 = - \vec \kappa + \vec k_\perp \,.
\end{eqnarray}

The set of eight vertex operators $\hat \Gamma_l$ corresponding to the $l$-th diagram 
in Fig.~\ref{fig:QQh-inclusive} reads
\begin{eqnarray}
&& \hat \Gamma_1= \hat{V}_2(\vec{k}_{13},\alpha_1) \hat{U}_2(\vec k_2,\alpha_2) \,, \nonumber \\
&& \hat \Gamma_2= \hat{U}_1(\vec k_1,\alpha_1) \hat{V}_1(\vec{k}_{23},\alpha_2) \,, \nonumber \\
&& \hat \Gamma_3= -\alpha_1 \hat{U}_1(\vec k_1,\alpha_1) \hat{V}_1(\vec{k}_{23}-\alpha_3 \vec{k}_\perp,\alpha_2) \,, \nonumber \\
&& \hat \Gamma_4= -\alpha_2 \hat{V}_2(\vec{k}_{13}-\alpha_3 \vec{k}_\perp,\alpha_1) \hat{U}_2(\vec k_2,\alpha_2) \,, \nonumber \\
&& \hat \Gamma_5= -\alpha_2\alpha_3 \hat{U}_1(\vec k_1-\vec{k}_\perp,\alpha_1) \hat{V}_1(\vec{k}_{23},\alpha_2) \,, \nonumber \\
&& \hat \Gamma_6= -\alpha_1\alpha_3 \hat{V}_2(\vec{k}_{13},\alpha_1) \hat{U}_2(\vec k_2-\vec{k}_\perp,\alpha_2) \,, \nonumber \\
&& \hat \Gamma_7= -\alpha_2\alpha_3 \hat{U}_1(\vec k_1-\alpha_1\vec{k}_\perp,\alpha_1) \hat{V}_1(\vec{k}_{23},\alpha_2) \,, \nonumber \\
&& \hat \Gamma_8=  \alpha_1\alpha_3 \hat{V}_2(\vec{k}_{13},\alpha_1) \hat{U}_2(\vec k_2-\alpha_2\vec{k}_\perp,\alpha_2) \,,
\end{eqnarray}
where
\begin{eqnarray}
&& \vec{k}_{13}\equiv \alpha_3 \vec{k}_1 - \alpha_1 \vec{k}_3 = \alpha_3 \vec \varkappa - \alpha \vec\kappa \,, \\ 
&& \vec{k}_{23}\equiv \alpha_3 \vec{k}_2 - \alpha_2 \vec{k}_3 = - \alpha_3 \vec \varkappa - \bar \alpha \vec\kappa \,,
\end{eqnarray}
and the $2\times2$ matrices $\hat U_{1,2}$ and $\hat V_{1,2}$ are given by
\begin{eqnarray} \nonumber
&& \hat{U}_1(\vec k_1,\alpha_1) = m_Q\,\vec{\sigma}\cdot \vec{e} + 
(1-2\alpha_1) (\vec{\sigma}\cdot \vec{n})(\vec{e}\cdot \vec{k}_1)+i(\vec{e}\times \vec{n})\cdot \vec{k}_1 \,, \\
&& \hat{U}_2(\vec k_2,\alpha_2) = m_Q\,\vec{\sigma}\cdot \vec{e} + 
(1-2\alpha_2) (\vec{\sigma}\cdot \vec{n})(\vec{e}\cdot \vec{k}_2)-i(\vec{e}\times \vec{n})\cdot \vec{k}_2 \,, \nonumber \\
&& \hat{V}_{1,2} = 2m_H\,\alpha_{2,1}\,(\alpha_{2,1}-\alpha_3) \,. \label{ops}
\end{eqnarray}
Here, $\vec{e}$ is the initial gluon $G_a$ polarisation vector, $\vec \sigma=\{\sigma^1,\sigma^2,\sigma^3\}$ is the vector of Pauli matrices $\sigma^a$, and
$\vec n$ is the unit vector in the direction of the corresponding particle momentum. The propagator functions $D_l$ which enter the denominator 
in Eq.~(\ref{amp-tot}) read:
\begin{eqnarray}  \nonumber
&& D_1=\Delta_0(\vec{k}_2)\Delta_2(\vec{k}_{13},\alpha_1,\alpha_2) \,, \\
&& D_2=\Delta_0(\vec{k}_1)\Delta_2(\vec{k}_{23},\alpha_2,\alpha_1) \,, \nonumber \\
&& D_3=\Delta_0(\vec{k}_1)\Delta_1(\vec{k}_1,\vec{k}_{23}-\alpha_3 \vec{k}_\perp,\alpha_1,\alpha_2,\alpha_3) \,, \nonumber \\
&& D_4=\Delta_0(\vec{k}_2)\Delta_1(\vec{k}_2,\vec{k}_{13}-\alpha_3 \vec{k}_\perp,\alpha_2,\alpha_1,\alpha_3) \,, \nonumber \\
&& D_5=\Delta_1(\vec{k}_1-\vec{k}_\perp,\vec{k}_{23},\alpha_1,\alpha_2,\alpha_3)\Delta_2(\vec{k}_{23},\alpha_2,\alpha_1) \,, \nonumber \\
&& D_6=\Delta_1(\vec{k}_2-\vec{k}_\perp,\vec{k}_{13},\alpha_2,\alpha_1,\alpha_3)\Delta_2(\vec{k}_{13},\alpha_1,\alpha_2) \,, \nonumber \\
&& D_7=\Delta_2(\vec{k}_{23},\alpha_2,\alpha_1)\Delta_1(\vec{k}_1-\alpha_1\vec{k}_\perp,\vec{k}_{23},\alpha_1,\alpha_2,\alpha_3) \,, \nonumber \\
&& D_8=\Delta_2(\vec{k}_{13},\alpha_1,\alpha_2)\Delta_1(\vec{k}_2-\alpha_2\vec{k}_\perp,\vec{k}_{13},\alpha_2,\alpha_1,\alpha_3) \,,
\end{eqnarray}
where 
\begin{eqnarray}
&& \Delta_0(\vec{k}_1) = m_Q^2 + \vec k_1^2 \,, \nonumber \\
&& \Delta_1(\vec{k}_1,\vec{k}_{23},\alpha_1,\alpha_2,\alpha_3) =
\alpha_2\alpha_3 \vec k_1^2 + \alpha_1 \vec k_{23}^2 + \alpha_1\bar\alpha_1\alpha_2 m_H^2 + \alpha_3(\alpha_2+\alpha_1\alpha_3)m_Q^2 \,, \nonumber \\
&& \Delta_2(\vec{k}_{13},\alpha_1,\alpha_2) = \alpha_3^2 m_Q^2 + \alpha_1 \bar \alpha_2 m_H^2 + \vec k_{13}^2 \,, \qquad
\bar \alpha_i = 1 - \alpha_i \,. \label{props}
\end{eqnarray}
It is worth to notice that $D_{1,2}$ functions are dependent on others since
\begin{eqnarray*}
[\Delta_0(\vec{k}_1)\Delta_2(\vec{k}_{23},\alpha_2,\alpha_1)]^{-1}&=&
\alpha_1 [\Delta_0(\vec{k}_1)\Delta_1(\vec{k}_1,\vec{k}_{23},\alpha_1,\alpha_2,\alpha_3)]^{-1} \\ &+&
\alpha_2\alpha_3 [\Delta_2(\vec{k}_{23},\alpha_2,\alpha_1)
\Delta_1(\vec{k}_1,\vec{k}_{23},\alpha_1,\alpha_2,\alpha_3)]^{-1}
\end{eqnarray*}
is satisfied. Together with the above relations, the latter one enables us to represent the total
Higgsstrahlung amplitude (\ref{amp-tot})
\begin{eqnarray} \label{amp-tot-1}
B^{\mu\bar\mu}_{a}=\frac{i\alpha_s\,m_Q}{\sqrt{4\pi}v}
\sum_{d=1}^{N_c^2-1} \frac{F^{(d)}_{Gp\to X}(\vec k_\perp,\{X\})}{\vec k_\perp^2+m_g^2}\, 
{\xi_Q^\mu}^\dagger  \Big\{ \tau_a\tau_d\, \hat T_1 + \tau_d\tau_a\, \hat T_2  \Big\} 
\tilde{\xi}_{\bar Q}^{\bar \mu} \,,
\end{eqnarray}
in terms of two independent helicity amplitudes
\begin{eqnarray} \nonumber
\hat T_1&=&\alpha_1\, \hat\nu_1(\vec{k}_1)\, \Big\{\hat\mu_1(\vec{k}_1,\vec{k}_{23})-\hat\mu_1(\vec{k}_1,\vec{k}_{23}-\alpha_3\vec{k}_\perp)\Big\}\, \\
 &+& \alpha_2\alpha_3\, \Big\{\hat\lambda_1(\vec{k}_1,\vec{k}_{23})-\hat\lambda_1(\vec{k}_1-\vec{k}_\perp,\vec{k}_{23})\Big\}\, 
 \hat\rho_1(\vec{k}_{23}) \nonumber \\
 &-& \alpha_1\alpha_3\, \hat\rho_2(\vec{k}_{13})\, \Big\{\hat\lambda_2(\vec{k}_2-\vec{k}_\perp,\vec{k}_{13})-
 \hat\lambda_2(\vec{k}_2-\alpha_2\vec{k}_\perp,\vec{k}_{13})\Big\} \,, \label{T1}\\ 
 \hat T_2&=&\alpha_2\, \Big\{\hat\mu_2(\vec{k}_2,\vec{k}_{13})-\hat\mu_2(\vec{k}_2,\vec{k}_{13}-\alpha_3\vec{k}_\perp)\Big\}\, 
 \hat\nu_2(\vec{k}_2) \nonumber \\
 &+& \alpha_1\alpha_3\, \hat\rho_2(\vec{k}_{13})\, \Big\{\hat\lambda_2(\vec{k}_2,\vec{k}_{13})-
 \hat\lambda_2(\vec{k}_2-\vec{k}_\perp,\vec{k}_{13})\Big\} \nonumber \\
 &-& \alpha_2\alpha_3\, \Big\{\hat\lambda_1(\vec{k}_1-\vec{k}_\perp,\vec{k}_{23})-
 \hat\lambda_1(\vec{k}_1-\alpha_1\vec{k}_\perp,\vec{k}_{23})\Big\}\, 
 \hat\rho_1(\vec{k}_{23}) \,, \label{T2}
\end{eqnarray}
where
\begin{eqnarray} \nonumber
&& \hat\nu_1(\vec{k}_1) = \frac{\hat{U}_1(\vec{k}_1,\alpha_1)}{\Delta_0(\vec{k}_1)} \,, \qquad 
     \hat \nu_2(\vec{k}_2) = \frac{\hat{U}_2(\vec{k}_2,\alpha_2)}{\Delta_0(\vec{k}_2)} \,, \\
&& \hat\rho_1(\vec{k}_{23}) = \frac{\hat{V}_1(\vec{k}_{23},\alpha_2)}
       {\Delta_2(\vec{k}_{23},\alpha_2,\alpha_1)} \,, \qquad 
     \hat \rho_2(\vec{k}_{13}) = \frac{\hat{V}_2(\vec{k}_{13},\alpha_1)}
       {\Delta_2(\vec{k}_{13},\alpha_1,\alpha_2)} \,, \nonumber \\  
&& \hat\mu_1(\vec{k}_1,\vec{k}_{23}) = \frac{\hat{V}_1(\vec{k}_{23},\alpha_2)}
       {\Delta_1(\vec{k}_1,\vec{k}_{23},\alpha_1,\alpha_2,\alpha_3)} \,, \quad 
     \hat \mu_2(\vec{k}_2,\vec{k}_{13}) = \frac{\hat{V}_2(\vec{k}_{13},\alpha_1)}
       {\Delta_1(\vec{k}_2,\vec{k}_{13},\alpha_2,\alpha_1,\alpha_3)} \,, \nonumber \\  
&& \hat\lambda_1(\vec{k}_1,\vec{k}_{23}) = \frac{\hat{U}_1(\vec{k}_1,\alpha_1)}
       {\Delta_1(\vec{k}_1,\vec{k}_{23},\alpha_1,\alpha_2,\alpha_3)} \,, \quad 
     \hat \lambda_2(\vec{k}_2,\vec{k}_{13}) = \frac{\hat{U}_2(\vec{k}_2,\alpha_2)}
       {\Delta_1(\vec{k}_2,\vec{k}_{13},\alpha_2,\alpha_1,\alpha_3)} \,.        
\end{eqnarray}
Clearly, the amplitudes $\hat T_{1,2}$ are related by a symmetry $\hat T_1 \leftrightarrow \hat T^\dagger_2$
w.r.t. $Q$ and $\bar Q$ interchange, i.e. $\vec{k}_1 \leftrightarrow \vec{k}_2$ and $\alpha_1 \leftrightarrow \alpha_2$.
Apparently, $\hat T_{1,2}$ vanish in the forward direction $\vec{k}_\perp \to 0$ which guarantees the infra-red
stability of the cross section.

The above expressions significantly simplify, if the longitudinal momentum fraction $\alpha_3$ carried by 
the emitted Higgs boson is small, i.e. $\alpha_3 \ll 1$, corresponding to the dominant configuration 
for the fluctuation of the projectile gluon, $G_a \to Q\bar{Q}+H$. Then the operators (\ref{ops}) are
\begin{eqnarray} \nonumber
&& \hat{U}_1(\vec k, \alpha) \simeq \hat{U}_2(-\vec k, \bar \alpha) \equiv \hat{U}(\vec k) = 
m_Q\,\vec{\sigma}\cdot \vec{e} + (1-2\alpha) (\vec{\sigma}\cdot \vec{n})(\vec{e}\cdot \vec{k}) + 
i(\vec{e}\times \vec{n})\cdot \vec{k} \,, \\
&& \hat{V}_1 \simeq \bar\alpha^2\,\hat{V}(\alpha_3/\bar\alpha)\,, \quad
\hat{V}_2 \simeq \alpha^2\,\hat{V}(\alpha_3/\alpha)\,, \quad \hat{V}(\gamma) \equiv 2(1-\gamma)\,m_H \,.\label{UV}
\end{eqnarray}
Insident quark transverse momenta become
\begin{eqnarray*}
&& \vec k_1 \simeq \vec \varkappa - \alpha (\vec \kappa - \vec k_\perp ) \,, \qquad  
\vec k_2 \simeq - \vec \varkappa - \bar\alpha (\vec \kappa - \vec k_\perp ) \,, \qquad
\vec k_3 \simeq \vec \kappa\,,
\end{eqnarray*}
and the propagators in Eq.~(\ref{props}) are
\begin{eqnarray} \nonumber
&& \Delta_0(\vec{k}_1) \simeq {\cal D}_1\big(\vec \varkappa - \alpha (\vec \kappa - \vec k_\perp ) \big)\,, \\
&& \Delta_0(\vec{k}_2) \simeq {\cal D}_1\big(\vec \varkappa + \bar\alpha (\vec \kappa - \vec k_\perp ) \big)\,, \quad 
{\cal D}_1(\vec k)\equiv \vec k^2 + m_Q^2 \,, \nonumber \\
&& \Delta_2(\vec{k}_{13},\alpha_1,\alpha_2) \simeq \alpha^2 {\cal D}_2\big(\vec\kappa - (\alpha_3/\alpha) \vec\varkappa,\alpha\big)\,, \nonumber  \\
&& \Delta_2(\vec{k}_{23},\alpha_2,\alpha_1) \simeq \bar\alpha^2 {\cal D}_2\big(\vec\kappa + 
(\alpha_3/\bar\alpha) \vec\varkappa,\bar\alpha\big)\,, \quad {\cal D}_2(\vec k,\alpha)\equiv \vec k^2 + \omega^2(\alpha) \,, \nonumber \\
&& \Delta_1(\vec{k}_1,\vec{k}_{23},\alpha_1,\alpha_2)\simeq 
\bar\alpha \big[\alpha_3 {\cal D}_1(\vec{k}_1) +\alpha\bar\alpha {\cal D}_2(\vec{k}_{23}/\bar\alpha,\bar\alpha)\big] \,, \nonumber \\
&& \Delta_1(\vec{k}_2,\vec{k}_{13},\alpha_2,\alpha_1)\simeq 
\alpha \big[\alpha_3 {\cal D}_1(\vec{k}_2) + \alpha\bar\alpha {\cal D}_2(\vec{k}_{13}/\alpha,\alpha)\big]\,,  \label{delta}
\end{eqnarray}
where
\begin{eqnarray}
\omega^2(\alpha) \simeq m_H^2 + \Big(\frac{\alpha_3}{\alpha}\Big)^2 m_Q^2 \,. \label{omega} 
\end{eqnarray}

The typical scales for incident transverse momenta are
\begin{eqnarray}
|\vec{k}_\perp| \sim m_g \ll |\vec{\kappa}|\,, \; |\vec{\varkappa}|\,, \qquad 
|\vec\kappa| \sim m_H \,, \qquad |\vec \varkappa| \sim m_Q \,, \label{hierar}
\end{eqnarray}
such that
\begin{eqnarray}
M^2\simeq M_{\bar QQ}^2+\vec\kappa^{\,2}+M_{-}^2\,,  \qquad 
M_{\bar QQ}^2\equiv \frac{m_Q^2 + \vec \varkappa^{\,2}}{\bar{\alpha}\alpha}\,, \qquad 
M_{-}^2\equiv \frac{m_H^2+\vec \kappa^{\,2}}{\alpha_3}\,. \label{invMass}
\end{eqnarray}

In the dominating configuration corresponding to $\alpha_3\ll \alpha \sim \bar\alpha$ 
asymptotics we have
\begin{eqnarray}
&& \Delta_1(\vec{k}_1,\vec{k}_{23},\alpha_1,\alpha_2) \simeq \alpha\bar \alpha^2( \vec \kappa^{\,2} + \tau^2 ) \,, \nonumber \\
&& \Delta_1(\vec{k}_2,\vec{k}_{13},\alpha_2,\alpha_1) \simeq \bar\alpha \alpha^2( \vec \kappa^{\,2} + \tau^2 ) \,,  \label{delta-1}
\end{eqnarray}
where
\begin{eqnarray}
&&\tau^2 \simeq m_H^2 + \alpha_3 M_{\bar QQ}^2\,. \label{tau}
\end{eqnarray}
In this asymptotics we finally get
\begin{eqnarray}
\hat T_1(\vec k_\perp,\vec \varkappa, \vec \kappa) &\simeq & 
\frac{\hat{U}\big(\vec \varkappa - \alpha (\vec \kappa - \vec k_\perp)\big)}
{{\cal D}_1\big(\vec \varkappa - \alpha (\vec \kappa - \vec k_\perp) \big)}\,  
\Big\{\frac{\hat{V}(\alpha_3/\bar\alpha)}{\tau^2 + \vec\kappa^2} - \frac{\hat{V}(\alpha_3/\bar\alpha)}
{\tau^2 + \big( \vec\kappa + (\alpha_3/\bar\alpha) \vec k_\perp \big)^2} \Big\} \,, \label{T1sS} \\
\hat T_2(\vec k_\perp,\vec \varkappa, \vec \kappa) &\simeq &    
\Big\{\frac{\hat{V}(\alpha_3/\alpha)}{\tau^2 + \vec\kappa^2} - \frac{\hat{V}(\alpha_3/\alpha)}
{\tau^2 + \big( \vec\kappa + (\alpha_3/\alpha) \vec k_\perp \big)^2} \Big\}\,
\frac{\hat{U}\big(\vec \varkappa + \bar\alpha (\vec \kappa - \vec k_\perp)\big)}
{{\cal D}_1\big(\vec \varkappa + \bar\alpha (\vec \kappa - \vec k_\perp) \big)} \,.\label{T2sS}
\end{eqnarray}

{\bf Acknowledgments}

Useful discussions with Valery Khoze, Antoni Szczurek and Gunnar
Ingelman are gratefully acknowledged. This study was partially
supported by Fondecyt (Chile) grants 1120920, 1130543 and 
1130549, and by ECOS-Conicyt grant No. C12E04. 
R. P. was partially supported by Swedish Research 
Council Grant No. 2013-4287.


\end{document}